\documentclass[oldversion]{aa} 
\usepackage{graphicx}
\usepackage{aalongtable}
\usepackage{txfonts}
\usepackage{rotating}
\usepackage{lscape}
\newcommand{\gsim}{\;\lower.6ex\hbox{$\sim$}\kern-7.75pt\raise.65ex\hbox{$>$}\;}
\newcommand{\lsim}{\;\lower.6ex\hbox{$\sim$}\kern-7.75pt\raise.65ex\hbox{$<$}\;}

\begin{document}

\title{Potassium abundances in multiple stellar populations of the globular
cluster NGC~4833\thanks{Based on observations collected at 
ESO telescopes under programmes 083.D-0208 and 095.D-0539.} 
 }

\author{
Eugenio Carretta\inst{1}
}

\authorrunning{Eugenio Carretta}
\titlerunning{Abundances of potassium in NGC~4833}

\offprints{E. Carretta, eugenio.carretta@inaf.it}

\institute{
INAF-Osservatorio di Astrofisica e Scienza dello Spazio di Bologna, Via Gobetti
 93/3, I-40129 Bologna, Italy}

\date{}

\abstract{NGC~4833 is a metal-poor Galactic globular cluster (GC) whose multiple
stellar populations present an extreme chemical composition. The Na-O
anti-correlation is quite extended, which is in agreement with the long tail on
the blue horizontal branch, and the large star-to-star variations in the [Mg/Fe]
ratio span more than 0.5 dex. Recently, significant excesses of Ca and Sc with
respect to field stars of a similar metallicity were also found, signaling the
production of species forged in H-burning at a very high temperature in the
polluters of the first generation in this cluster. Since an enhancement of
potassium is also expected under these conditions, we tested this scenario by
analysing intermediate resolution spectra of 59 cluster stars including the
K~{\sc i} resonance line at 7698.98~\AA. We found a wide spread of K abundances,
anti-correlated to Mg and O abundances, as previously also observed in NGC~2808.
The abundances of K are found to be correlated to those of Na, Ca, and Sc.
Overall, this chemical pattern confirms that NGC~4833 is one of the relatively
few GCs where the self-enrichment from first generation polluters occurred at
such high temperatures that proton-capture reactions were able to proceed up to
heavier species such as K and possibly Ca. The spread in K observed in GCs
appears to be a function of a linear combination of cluster total luminosity and
metallicity, as other chemical signatures of multiple stellar populations in
GCs.
}

\maketitle

\section{Introduction}

Multiple stellar populations in Galactic globular clusters (GCs) are commonly
recognised as the fossil records of a chemical self-enrichment at early epochs
in the cluster lifetime. The enhancements and depletions in light element
abundances currently observed in most GC stars very likely originated in already
evolved massive stars of the first generation (FG) that formed in the cluster.
This stems from the observations, covering several decades nowadays (see Smith
1987, Kraft 1994, Gratton et al. 2004, 2012), that  depleted elements are
anti-correlated to enhanced species and correlated to other depleted species;
this is an overall chemical pattern that is well explained by proton-capture
reactions occurring in high temperature H-burning (e.g. Denisenkov and
Denisenkova 1989).

However, determining what exactly these FG polluters  were is still
controversial.  The imperfect match between observations and models introduces
uncertainties on the probable sites for this peculiar nucleosynthesis in GCs
(e.g. the recent review by Bastian and Lardo 2018, Gratton et al. 2019). For a
more unambigous answer to this issue, it is important to sample as many GCs as
possible and as many abundances of the different atomic species involved.
Lighter elements such as C, N, O, and Na (forged at a relatively moderate
temperature in H-burning and the first species used to tag multiple populations
in GCs) are not enough. Heavier species must be analysed to sample the regime(s)
where the addition of protons requires higher and higher temperatures to
overcome the Coulomb barrier. In turn, different temperature regimes can be
translated  through stellar evolutionary models into mass regimes, providing
useful constraints to define the nature of the FG polluters.

A good example is represented by the K-Mg anti-correlation discovered in
NGC~2419 (Cohen and Kirby 2012, Mucciarelli et al. 2012). In the context of
multiple populations, Ventura et al. (2012) proposed that at the low metallicity
of this cluster, the H-burning occurred to such a high temperature that the
consumption of Mg was accompanied not only by the usual Al production, but also
by the synthesis of heavier elements such as K, and even possibly Ca, from
proton-capture on Ar nuclei. They proposed super asymptotic giant branch (S-AGB)
stars as polluters, which were later confirmed by Iliadis et al. (2016) as viable
contributors, provided that models are calibrated in such a way to give the
highest possible temperatures.

\object{NGC~4833} is a good target where to search for the chemical fossil records 
left from such a regime. It is a massive (total absolute magnitude
$M_V=-8.17$, Harris 1996, 2010 online edition) and metal-poor ([Fe/H]$=-2.02$
dex, Carretta et al. 2014) GC. This is exactly the right
combination of global parameters for generating both large depletions in O
(Carretta et al. 2009a) and high levels of Al production (Carretta et al. 2009b;
see also Pancino et al. 2017, M\'esz\'aros et al. 2020), as well as significant
excesses of Ca and Sc from proton-capture reactions (Carretta and Bragaglia
2021, their fig.9).
Unlike the case of the Mg-Al anti-correlation, observed in a good fraction of 
GCs, the very  high temperature regime for hot H-burning, with alterations
potentially affecting the K, Ca, and Sc group, is still poorly sampled. Any new
addition would represent a precious constraint to properly tune stellar models.

Already the first extensive abundance analysis in NGC~4833 revealed that the
chemical composition of the multiple populations in this cluster is extreme
(Carretta et al. 2014). The Na-O anti-correlation is found to reach a long
extension, as predicted by the robust relation between the extent of this
signature and the hottest point along the horizontal branch (HB: Carretta et al.
2007a). The long blue tail of the HB clearly stands out in NGC~4833 (see e.g.
the HST snapshot survey by Piotto et al. 2002). Beside the long extent of the
Na-O anti-correlation, an unusually wide spread in the [Mg/Fe] abundance ratio
was detected; this was later independently confirmed by Roederer and Thompson
(2015), although with a smaller sample.

In Carretta et al. (2014), a correlation between a Ca and Sc abundance was also
observed, but it was overlooked because it was erroneously considered not to be
significant. Nevertheless, in the diagnostic plot of [Ca/Mg] versus [Ca/H]
employed to evaluate the possible relation between Mg and heavier elements, we
found that NGC~4833 stands out in the [Ca/Mg] distribution with respect to many
GC stars, together with  NGC~2808, NGC~5139 ($\omega$ Cen), NGC~6715 (M~54), and
NGC~7078 (M~15). This behaviour points out that the observed large depletions in
Mg are likely accompanied by excesses in Ca abundances.

A confirmation for these findings was recently found in a census of 
Mg, Ca, and Sc abundances in a large sample of GCs (Carretta and Bragaglia
2021). Using an unpolluted sample of field stars as a reference, significant 
excesses of Ca and Sc were detected in a number of GCs, almost all the
ones picked up by the diagnostic plot in Carretta et al. (2014), NGC~4833
included. 

Ideally, it would also be important to check the chemical pattern found
for elements such as P and S, which are intermediate steps to Ca and Sc in the
proton-capture sequence. Unfortunately, the few pioneering studies are  still
limited to a handful of stars in the GCs closer to the Sun, none of which are
among the extreme cases listed above. Phosphorus abundances were derived in
HB stars of NGC~6397 and NGC~6752 (Hubrig et al. 2009), whereas sulphur was
observed in stars of NGC~6752 and NGC~104 (47~Tuc) by Sbordone et al. (2009), in
one star of NGC~6397 by Koch and Caffau (2011), and in NGC~6121 (M~4), NGC~6656
(M~22), and NGC~7099 (M~30) by Kacharov et al. (2015). Interestingly, in 47~Tuc
S, abundances seem to be correlated to the Na abundances, suggesting a possible
origin through proton-capture reactions.

To better assess the right framework in which to interpret these data in the
context of multiple populations in GCs, it would be advantageous to also
directly check the other element involved, potassium. Abundance analysis of K in
GC stars is, however, scanty. Apart from the notable case of NGC~2419 cited
above, only about a dozen GCs have been scrutinised for K (Takeda et al. 2009,
Roederer et al. 2011, Carretta et al. 2013, Roederer and Thompson 2015,
Mucciarelli et al. 2015, 2017, M\'esz\'aros et al. 2020). Only in NGC~2808
(Mucciarelli et al. 2015) and $\omega$ Cen  (M\'esz\'aros et al. 2020)
significant star-to-star abundance variations in [K/Fe] were detected, albeit
the intrinsic spread is much smaller than the huge dispersion observed in
NGC~2419. 

Moreover, the samples are often limited to a few stars in each GC, and this 
precludes any firm conclusion from being drawn. The analysis of infrared data from
APOGEE (M\'esz\'aros et al. 2020) revealed only two K-rich stars in NGC~2808, due
to the severe selection criteria imposed on abundance quality, and in NGC~1904
the data dispersion in Mg is unfortunately comparable to the uncertainties, so
that in this GC the existence of a K-Mg anti-correlation is at least weak or
questionable.

In NGC~4833 Roederer and Thompson (2015) confirmed a large dispersion in Mg
abundances, with a bi-modal distribution of [Mg/Fe] ratios from a sample of 15
giants. They also derived K abundances for all the stars; however, apparently
after dropping the star with the highest [K/Fe] ratio (star 4-224 seems to have 
been excluded from both their plots and averages), they concluded that the
evidence for a K-Mg anti-correlation was not compelling in this cluster. The same
conclusion was reached for variations in Ca and Sc.

Fortunately, NGC~4833 is one of the few GCs whose data in the ESO archive do
include spectra in the region where the transitions of the resonance doublet of
K~{\sc i} at 7664 and 7698~\AA\ fall. This occurrence allowed us to derive the
[K/Fe] ratios for a large number of cluster stars whose abundances of other
proton-capture elements were already homogeneously determined in Carretta et al.
(2014). The present analysis reveals a significant dispersion in K abundances
and a pattern of correlation and anti-correlations with other light species well
explained by the outcome of very high temperature H-burning in multiple stellar
populations in GCs.

In Section 2 we describe the data and the analysis. In Section 3 we discuss our
results, and in Section 4 we frame our new findings for NGC~4833 in the context
of what we know for the behaviour of heavy proton-capture elements in GCs. A
summary of the main points is given in Section 5.

\begin{figure}
\centering
\includegraphics[scale=0.40]{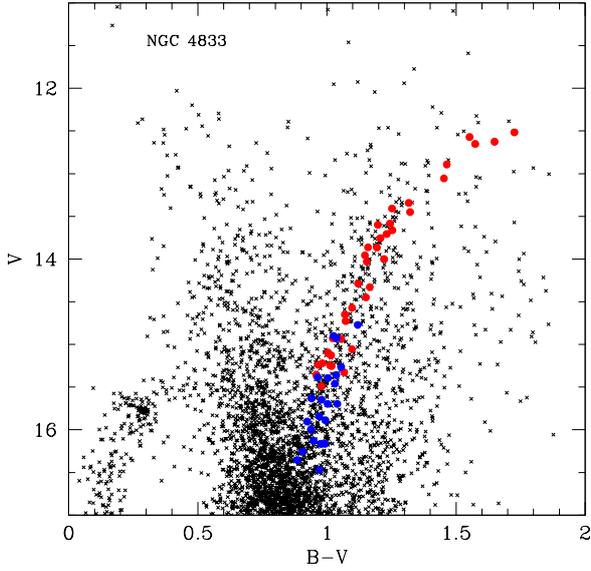}
\caption{Colour-magnitude diagram $V,B-V$ from the photometric catalogue in
Carretta et al. (2014: small crosses). Stars selected for the present study are
represented as larger filled circles. Blue symbols indicate stars with lower S/N
spectra (see text).}
\label{f:fig1}
\end{figure}

\section{Observations and abundance analysis}

 From the ESO archive of advanced data products, we downloaded the 
one-dimensional, wavelength-calibrated spectra of stars in NGC~4833 including
the resonance doublet of K~{\sc i}. The spectra were acquired on 30 May 2015 in
two exposures, each 1200 sec long, under the ESO programme 095.D-0539 (P.I.
Mucciarelli), using the GIRAFFE high resolution setup HR18 (R=19,000), covering
the spectral range from about 7460 to about 7883~\AA.

\begin{figure}
\centering
\includegraphics[scale=0.40]{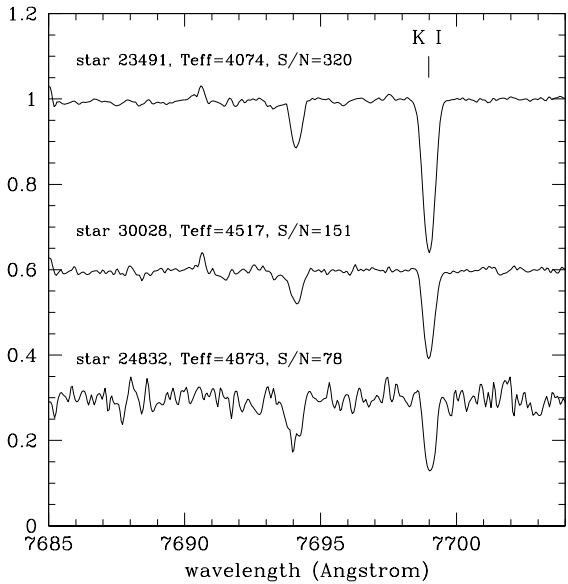}
\caption{Example of cleaned spectra in three ranges of S/N.}
\label{f:fig2}
\end{figure}

We then selected the spectra for a subset of 59 stars whose atmospheric
parameters, metallicities, and element abundances were derived in Carretta et al.
(2014) using procedures homogeneously adopted for all GCs in our FLAMES survey
(see Carretta et al. 2006 for details). The survey also includes NGC~2808
(Carretta 2015), the only other GC to date besides NGC~2419 where a clear K-Mg
anti-correlation was detected (Mucciarelli et al. 2015).

The position of the selected stars is shown on the $V,B-V$ colour-magnitude
diagram in Fig.~\ref{f:fig1} from the photometric catalogue described in
Carretta et al. (2014). In Table~\ref{t:tab1} we list star identification,
coordinates, and atmospheric parameters from Carretta et al. (2014).

\setcounter{table}{0}
\begin{table*}
\centering
\caption{Coordinates and atmospheric parameters of selected stars in NGC~4833.}
\begin{tabular}{ccccccr}
\hline
star  & RA         & DEC          & T$_{\rm eff}$ & $\log g$ & [A/H] & S/N \\   
\hline
22810 & 194.773166 & $-$70.949972 & 4893 & 2.17 & $-$2.04 &    105  \\ 
23306 & 194.707041 & $-$70.921388 & 4914 & 2.20 & $-$2.03 &     65  \\ 
23437 & 194.728583 & $-$70.916388 & 4767 & 1.89 & $-$2.06 &    110  \\ 
23491 & 194.771666 & $-$70.914944 & 4074 & 0.45 & $-$2.04 &    320  \\ 
23518 & 194.802833 & $-$70.913888 & 4867 & 2.12 & $-$2.02 &     86  \\ 
24063 & 194.725416 & $-$70.896333 & 4352 & 1.04 & $-$2.05 &    219  \\ 
24252 & 194.816999 & $-$70.891361 & 4432 & 1.20 & $-$1.99 &    224  \\ 
24339 & 194.709749 & $-$70.889249 & 4992 & 2.38 & $-$2.01 &     60  \\ 
24383 & 194.783791 & $-$70.888055 & 4766 & 1.91 & $-$2.02 &     89  \\ 
24515 & 194.817166 & $-$70.884499 & 4748 & 1.86 & $-$2.02 &    131  \\ 
24575 & 194.658749 & $-$70.883027 & 4845 & 2.07 & $-$2.00 &    115  \\ 
24695 & 194.742958 & $-$70.880111 & 4746 & 1.86 & $-$2.03 &     89  \\ 
24727 & 194.809958 & $-$70.879305 & 4506 & 1.36 & $-$2.07 &    232  \\ 
24832 & 194.803499 & $-$70.876638 & 4873 & 2.13 & $-$2.02 &     78  \\ 
24877 & 194.730291 & $-$70.875638 & 4888 & 2.17 & $-$2.01 &     76  \\ 
26174 & 194.807333 & $-$70.841444 & 4931 & 2.25 & $-$2.07 &     67  \\ 
26207 & 194.760708 & $-$70.840083 & 4872 & 2.13 & $-$2.04 &    104  \\ 
28662 & 194.871541 & $-$70.963138 & 5142 & 2.70 & $-$2.03 &     47  \\ 
29105 & 195.138166 & $-$70.941777 & 4130 & 0.57 & $-$2.07 &    299  \\ 
29160 & 195.092083 & $-$70.940166 & 4665 & 1.70 & $-$2.05 &    112  \\ 
29192 & 194.926333 & $-$70.939138 & 4909 & 2.22 & $-$2.06 &     68  \\ 
29240 & 194.946666 & $-$70.937194 & 4539 & 1.43 & $-$2.07 &    130  \\ 
29414 & 195.180791 & $-$70.930638 & 4740 & 1.86 & $-$2.03 &     92  \\ 
29466 & 194.928124 & $-$70.929777 & 4268 & 0.86 & $-$2.05 &    223  \\ 
29711 & 194.873749 & $-$70.923055 & 5059 & 2.52 & $-$2.03 &     49  \\ 
29761 & 194.987083 & $-$70.922083 & 5075 & 2.56 & $-$2.06 &     54  \\ 
29786 & 195.097791 & $-$70.921499 & 4980 & 2.38 & $-$2.05 &     77  \\ 
30028 & 194.894624 & $-$70.916666 & 4517 & 1.39 & $-$2.05 &    151  \\ 
30287 & 194.895374 & $-$70.911888 & 4851 & 2.09 & $-$2.04 &     69  \\ 
30312 & 194.909166 & $-$70.911499 & 4499 & 1.36 & $-$2.01 &    234  \\ 
30424 & 194.835041 & $-$70.909638 & 4858 & 2.10 & $-$2.02 &    120  \\ 
30715 & 195.007541 & $-$70.905166 & 5129 & 2.70 & $-$2.01 &     45  \\ 
30828 & 194.852749 & $-$70.903694 & 4993 & 2.38 & $-$2.03 &     61  \\ 
30846 & 194.933624 & $-$70.903527 & 4157 & 0.63 & $-$2.05 &    287  \\ 
31370 & 194.872999 & $-$70.897527 & 4384 & 1.11 & $-$2.02 &    300  \\ 
31728 & 194.894708 & $-$70.893805 & 4952 & 2.30 & $-$2.01 &     62  \\ 
31836 & 194.912583 & $-$70.892722 & 4623 & 1.62 & $-$2.04 &    129  \\ 
31929 & 194.828541 & $-$70.891944 & 4618 & 1.60 & $-$2.04 &    120  \\ 
32359 & 194.928666 & $-$70.888166 & 4702 & 1.78 & $-$2.02 &    102  \\ 
33027 & 194.943499 & $-$70.882833 & 4725 & 1.83 & $-$2.06 &    141  \\ 
33183 & 195.017541 & $-$70.881638 & 4698 & 1.78 & $-$2.04 &    102  \\ 
33459 & 194.948374 & $-$70.879638 & 4457 & 1.26 & $-$2.07 &    200  \\ 
33956 & 195.056791 & $-$70.875638 & 5082 & 2.59 & $-$2.02 &     50  \\ 
34384 & 195.043083 & $-$70.872361 & 4787 & 1.96 & $-$2.09 &    114  \\ 
34613 & 194.823541 & $-$70.870749 & 4219 & 0.76 & $-$2.01 &    253  \\ 
34785 & 194.986333 & $-$70.869388 & 4149 & 0.62 & $-$2.02 &    274  \\ 
34907 & 195.066999 & $-$70.868416 & 4834 & 2.06 & $-$2.05 &    122  \\ 
35066 & 194.958833 & $-$70.867138 & 4551 & 1.46 & $-$2.05 &    133  \\ 
35302 & 194.821999 & $-$70.865277 & 4852 & 2.09 & $-$2.04 &    103  \\ 
35695 & 195.041124 & $-$70.861638 & 4884 & 2.18 & $-$2.02 &    111  \\ 
35808 & 194.865374 & $-$70.860555 & 5023 & 2.45 & $-$2.02 &     56  \\ 
35994 & 194.991916 & $-$70.858888 & 5146 & 2.73 & $-$2.03 &     49  \\ 
36201 & 195.058499 & $-$70.856749 & 5031 & 2.48 & $-$2.01 &     52  \\ 
36391 & 194.895666 & $-$70.855111 & 4437 & 1.22 & $-$2.04 &    255  \\ 
36454 & 194.947333 & $-$70.854472 & 4454 & 1.26 & $-$2.01 &    238  \\ 
36689 & 194.909083 & $-$70.852083 & 4829 & 2.05 & $-$2.02 &    148  \\ 
36716 & 194.922999 & $-$70.851833 & 4391 & 1.13 & $-$2.09 &    276  \\ 
37498 & 194.878666 & $-$70.841749 & 4535 & 1.43 & $-$2.02 &    130  \\ 
38029 & 194.866041 & $-$70.831916 & 4876 & 2.16 & $-$2.01 &    114  \\ 

\hline
\end{tabular}
\label{t:tab1}
\end{table*}

For each exposure, the spectra were sky-subtracted and the K~{\sc i} lines were
carefully examined. The line at 7664.91~\AA\ is heavily contaminated by strong
telluric lines, whereas the 7698.98~\AA\ line is only affected by contamination
from a weaker line. The observed spectra in each exposure were then cleaned by
dividing for a synthetic spectrum of the telluric lines over the region
7681-7710~\AA, following the procedure described for the [O~{\sc i}] forbidden
line in Carretta et al. (2006). Finally, for each star a co-added spectrum was
obtained from the cleaned spectra. The signal-to-noise ratio (S/N) of each
combined spectrum, estimated in a  region close to the K~{\sc i} line, is
reported in column (7) of Table~\ref{t:tab1}. Examples of co-added spectra are
shown in Fig.~\ref{f:fig2}.

Abundances of K were derived from the equivalent widths (EWs) of the K~{\sc i}
7698.98~\AA\ line measured on the cleaned spectra with the package ROSA (Gratton
1988). For the abundance analysis, we adopted the atmospheric parameters derived
in Carretta et al. (2014) with the exception of microturbulent velocity, $v_t$.
When using $v_t$ values obtained using weak Fe lines together with strong lines
such as the K~{\sc i} ones, being more sensitive to velocity fields in the
stellar atmospheres, a clear trend of abundances as a function of the $v_t$
appears, which is a well known effect (e.g. Carretta et al. 2014, Mucciarelli et
al. 2015). To alleviate this problem, we adopted the same approach used in
Carretta et al. (2014) for the strong Ba lines. We adopted the values of $v_t$ 
obtained from a relation as a function of the surface gravity derived by Worley
et al. (2013) for giants in the metal-poor GC M~15, whose metallicity is
comparable to that of NGC~4833. This relation was found to be efficient in
removing the trend of Ba abundances and it also works well for K abundances, as
in the present work. The same approach was adopted by Mucciarelli et al. (2015)
for NGC~2808, using a slightly different relation from Kirby et al. (2009).

As in Carretta et al. (2013), we applied corrections for departure from the
Local Thermodynamic Equilibrium (LTE) to
the abundances of K, using a multivariate interpolation as a function of
the temperature, gravity, metallicity, and the EW of the K~{\sc i} line
from the set of models by Takeda et al. 
(2002)\footnote{$http://optik2.mtk.nao.ac.jp/\sim takeda/potassium\_nonlte$}.
Due to the similar range of parameters for the present sample of stars, these
corrections are essentially a constant shift of the K abundances downwards by
about -0.6 dex, with a root mean square (r.m.s.) scatter of only 0.032 dex which
has a negligible impact on the star-to-star internal errors.

Internal errors were derived using sensitivities of abundances to variations in
the atmospheric parameters and the internal uncertainties in each parameter as
estimated in Carretta et al. (2014). The resulting star-to-star errors were then
summed in quadrature to uncertainties in the EW measurements, estimated by the formula from
Cayrel (1988). This contribution depends on the S/N, so that the
total error budget is 0.061 dex, 0.070 dex, and 0.095 dex in the ranges
S/N$>200$, 100$<$S/N$<$200, and S/N$<100$, respectively. In the end, we assumed
the average 0.08 dex as the average internal error associated with the K
abundances. We note that the systematic uncertainty related to the
trend of abundances as a function of $v_t$, if not properly treated, would
result in spurious changes in the [K/Fe] ratios potentially much larger than the
derived internal errors.
The derived [K/Fe] ratios, corrected for non-LTE, are listed in
Table~\ref{t:tab2}, together with the abundances of light elements derived for
programme stars in Carretta et al. (2014), for an easier comparison.

\setcounter{table}{1}
\begin{table*}
\centering
\caption{Abundances of light elements for the programme stars.}
\begin{tabular}{crrrrrrr}
\hline
star  &    [O/Fe]  &  [Na/Fe]  &  [Mg/Fe] &  [Si/Fe] &  [Ca/Fe] &  [Sc/Fe]  &  [K/Fe]    \\
\hline
22810 &    +0.227  &  +0.532   &  +0.336  &  +0.441  &  +0.356  &$-$0.044   &  +0.046      \\ 
23306 &    +0.623  &  +0.357   &  +0.592  &  +0.393  &  +0.353  &$-$0.030   &  +0.113      \\ 
23437 &            &  +0.358   &  +0.393  &  +0.511  &  +0.312  &$-$0.059   &  +0.092      \\ 
23491 &  $-$0.013  &  +0.805   &  +0.255  &  +0.453  &  +0.341  &  +0.003   &  +0.256      \\ 
23518 &    +0.546  &           &  +0.469  &  +0.412  &  +0.384  &$-$0.035   &  +0.323      \\ 
24063 &  $-$0.199  &  +0.890   &          &  +0.403  &  +0.365  &$-$0.050   &  +0.125      \\ 
24252 &    +0.482  &  +0.063   &  +0.575  &  +0.454  &  +0.352  &$-$0.047   &  +0.144      \\ 
24339 &    +0.507  &           &          &  +0.449  &  +0.341  &$-$0.059   &  +0.167      \\ 
24383 &            &           &          &  +0.414  &  +0.354  &$-$0.053   &  +0.073      \\ 
24515 &    +0.038  &  +0.584   &  +0.103  &  +0.507  &  +0.356  &$-$0.027   &  +0.296      \\ 
24575 &            &  +0.437   &  +0.323  &  +0.440  &  +0.320  &$-$0.050   &$-$0.054      \\ 
24695 &    +0.112  &  +0.483   &  +0.284  &  +0.472  &  +0.351  &$-$0.036   &  +0.338      \\ 
24727 &    +0.398  &  +0.171   &  +0.413  &  +0.365  &  +0.316  &$-$0.030   &  +0.029      \\ 
24832 &    +0.060  &  +0.531   &  +0.426  &  +0.425  &  +0.373  &$-$0.050   &  +0.224      \\ 
24877 &    +0.147  &           &          &          &  +0.366  &$-$0.015   &  +0.205      \\ 
26174 &            &           &          &          &  +0.347  &$-$0.020   &  +0.125      \\ 
26207 &    +0.045  &  +0.498   &  +0.244  &  +0.427  &  +0.348  &$-$0.010   &  +0.152      \\ 
28662 &            &  +0.573   &  +0.308  &  +0.518  &  +0.330  &$-$0.019   &  +0.329      \\ 
29105 &    +0.422  &  +0.479   &  +0.439  &  +0.485  &  +0.356  &$-$0.049   &$-$0.035      \\ 
29160 &    +0.254  &  +0.068   &  +0.507  &  +0.401  &  +0.366  &$-$0.026   &  +0.052      \\ 
29192 &            &  +0.546   &  +0.223  &  +0.495  &  +0.342  &$-$0.060   &  +0.165      \\ 
29240 &  $-$0.137  &  +0.711   &  +0.166  &  +0.428  &  +0.355  &$-$0.030   &  +0.061      \\ 
29414 &    +0.020  &  +0.613   &          &  +0.470  &  +0.357  &$-$0.032   &  +0.103      \\ 
29466 &    +0.209  &  +0.872   &  +0.429  &  +0.476  &  +0.351  &  +0.002   &$-$0.084      \\ 
29711 &            &  +0.817   &          &  +0.458  &  +0.349  &  +0.000   &$-$0.045      \\ 
29761 &            &           &          &          &  +0.388  &$-$0.009   &  +0.086      \\ 
29786 &    +0.340  &  +0.393   &  +0.407  &  +0.457  &  +0.365  &$-$0.053   &$-$0.021      \\ 
30028 &    +0.295  &  +0.371   &  +0.356  &  +0.437  &  +0.358  &$-$0.002   &$-$0.152      \\ 
30287 &    +0.548  &           &          &  +0.485  &  +0.360  &$-$0.039   &$-$0.002      \\ 
30312 &    +0.426  &  +0.143   &  +0.490  &  +0.420  &  +0.333  &$-$0.021   &  +0.025      \\ 
30424 &    +0.419  &  +0.412   &  +0.322  &  +0.429  &  +0.352  &$-$0.066   &  +0.182      \\ 
30715 &    +0.605  &  +0.175   &          &  +0.450  &  +0.369  &$-$0.043   &  +0.175      \\ 
30828 &    +0.282  &  +0.545   &          &  +0.429  &  +0.370  &$-$0.036   &  +0.046      \\ 
30846 &    +0.124  &  +0.947   &          &  +0.398  &  +0.334  &$-$0.059   &$-$0.020      \\ 
31370 &    +0.497  &  +0.235   &  +0.461  &  +0.465  &  +0.339  &$-$0.037   &  +0.109      \\ 
31728 &    +0.284  &  +0.207   &          &  +0.359  &  +0.332  &$-$0.036   &  +0.134      \\ 
31836 &    +0.351  &           &          &  +0.490  &  +0.334  &$-$0.061   &  +0.023      \\ 
31929 &    +0.425  &$-$0.098   &  +0.478  &  +0.446  &  +0.361  &$-$0.042   &  +0.043      \\ 
32359 &    +0.013  &  +0.579   &  +0.216  &  +0.527  &  +0.353  &$-$0.040   &  +0.070      \\ 
33027 &    +0.273  &  +0.293   &  +0.491  &  +0.457  &  +0.357  &$-$0.032   &  +0.103      \\ 
33183 &            &  +0.644   &  +0.197  &  +0.493  &  +0.361  &$-$0.017   &  +0.117      \\ 
33459 &  $-$0.381  &  +0.787   &  +0.191  &  +0.567  &  +0.348  &$-$0.027   &  +0.050      \\ 
33956 &            &           &          &  +0.449  &  +0.353  &$-$0.078   &  +0.203      \\ 
34384 &    +0.401  &           &          &  +0.479  &  +0.336  &$-$0.059   &$-$0.383      \\ 
34613 &  $-$0.585  &  +0.895   &$-$0.004  &  +0.634  &  +0.363  &$-$0.014   &  +0.053      \\ 
34785 &    +0.415  &  +0.123   &  +0.410  &  +0.455  &  +0.340  &$-$0.085   &$-$0.203      \\ 
34907 &    +0.514  &           &          &  +0.489  &  +0.342  &$-$0.025   &$-$0.132      \\ 
35066 &    +0.382  &  +0.597   &  +0.457  &  +0.461  &  +0.356  &$-$0.029   &  +0.077      \\ 
35302 &    +0.524  &  +0.031   &  +0.518  &  +0.482  &  +0.334  &$-$0.044   &  +0.064      \\ 
35695 &    +0.327  &           &          &  +0.503  &  +0.342  &$-$0.026   &  +0.310      \\ 
35808 &            &           &  +0.244  &  +0.486  &  +0.376  &$-$0.046   &  +0.309      \\ 
35994 &            &  +0.406   &          &          &  +0.325  &$-$0.023   &$-$0.057      \\ 
36201 &            &           &          &  +0.467  &  +0.363  &  +0.015   &$-$0.007      \\ 
36391 &  $-$0.405  &  +0.775   &  +0.021  &  +0.610  &  +0.361  &$-$0.042   &  +0.114      \\ 
36454 &    +0.378  &  +0.121   &  +0.450  &  +0.378  &  +0.355  &$-$0.060   &$-$0.067      \\ 
36689 &    +0.622  &  +0.022   &  +0.713  &  +0.440  &  +0.358  &$-$0.093   &$-$0.141      \\ 
36716 &    +0.053  &  +0.779   &  +0.293  &  +0.506  &  +0.374  &$-$0.008   &  +0.012      \\ 
37498 &  $-$0.010  &  +0.678   &  +0.212  &  +0.437  &  +0.355  &$-$0.048   &  +0.104      \\ 
38029 &    +0.454  &           &          &  +0.500  &  +0.334  &$-$0.066   &  +0.159      \\ 

\hline
\end{tabular}
\label{t:tab2}
\end{table*}

\section{Results}

Abundances of K were obtained for all of the 59 programme stars. In the
following, we explore the relation of K to the other proton-capture elements
analysed in Carretta et al. (2014) and the comparison with the previous analysis
by Roederer and Thompson (2015). The properties of NGC~4388 in the general
context of multiple stellar populations in GCs are discussed in the next section.

\subsection{Potassium and other light elements in NGC~4833}

\begin{figure}
\centering
\includegraphics[scale=0.40]{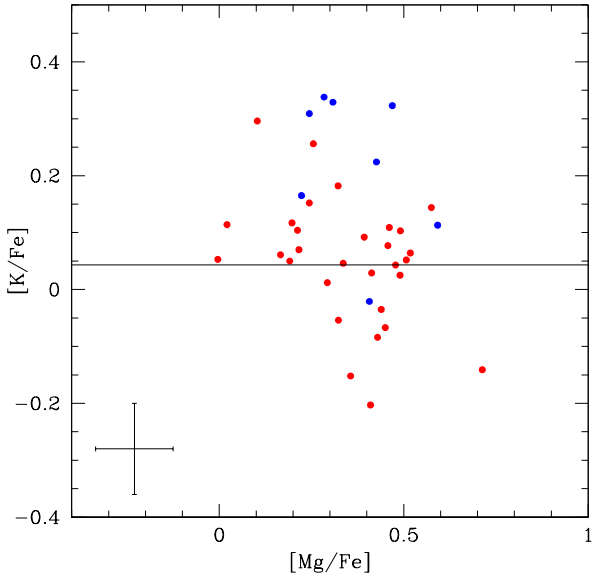}
\caption{Abundance ratios [K/Fe] from the present analysis as a function of
[Mg/Fe] ratios from Carretta et al. (2014). Blue points indicate stars whose
co-added spectra are of a lower quality (S/N$<100$). The horizontal line
is at the average value of [K/Fe] derived from high S/N spectra. Internal error
bars are also shown.}
\label{f:fig3}
\end{figure}

Our main result is summarised in Fig.~\ref{f:fig3}, where we plotted the
abundances of K from the present work as a function of the [Mg/Fe] ratios from
Carretta et al. (2014). In NGC~4833 we observe a large range of K abundances,
about 0.5 dex from peak to peak, which are comparable to the wide spread
detected for Mg abundances. The mean value of [K/Fe] for programme stars is
0.078 dex ($\sigma=0.137$ dex, 59 stars). Considering only stars with high S/N
spectra, the mean value slightly decreases to 0.043 dex, but the scatter remains
almost unvaried ($\sigma=0.136$ dex, 37 stars). The large scatter seems to be an
intrinsic feature: No significant trend is observed as a function of the
atmospheric parameters (effective temperature, microturbulent velocity). The
star-to-star variation in [K/Fe] is observed over four full magnitudes along the
red giant branch (RGB) in NGC~4833.

In Fig.~\ref{f:fig3} we plotted stars whose combined spectra have S/N$<100$  as
blue symbols: Almost all of them lie at the upper envelope of the K
distribution. This occurrence may suggest that measurements of K in such a low
metallicity GC could be affected by some additional residual noise when spectra
of a lower quality are used. We conservatively distinguish these stars with a
different colour in Fig.~\ref{f:fig3} and in the next figures, although our
conclusions are not affected by this consideration much.

A clear anti-correlation is observed in Fig.~\ref{f:fig3}: The higher abundances
of K are found for stars with the lower Mg abundances, and vice versa. To better
quantify these findings, we considered the average value [Mg/Fe]=0.27 dex from
the high resolution UVES spectra in Carretta et al. (2014). When the sample is
split at this value, the mean values of K abundances in the two groups are
[K/Fe]=+0.146 dex ($\sigma=0.093$ dex, 12 stars) and  [K/Fe]=+0.060 dex
($\sigma=0.142$ dex, 26 stars) for the Mg-poor and Mg-rich sub-samples,
respectively. With the Student's test, we checked the null hypothesis that these
two components are extracted from a distribution having the same mean [K/Fe]
ratio. This hypothesis can be rejected at a high level of confidence (t=2.266,
36 d.o.f., two-tail probability $p=0.03$).

\begin{figure}
\centering
\includegraphics[scale=0.40]{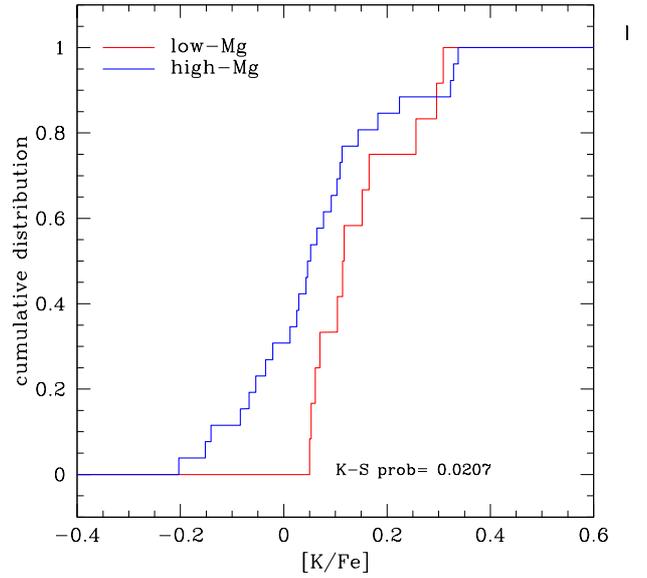}
\caption{Cumulative distributions of [K/Fe] abundance ratios
for the stars with [Mg/Fe] higher and lower than 0.27 dex. The
probability for the Kolmogorov-Smirnov test is also listed.}
\label{f:fig4}
\end{figure}

\begin{figure*}
\centering
\includegraphics[scale=0.30]{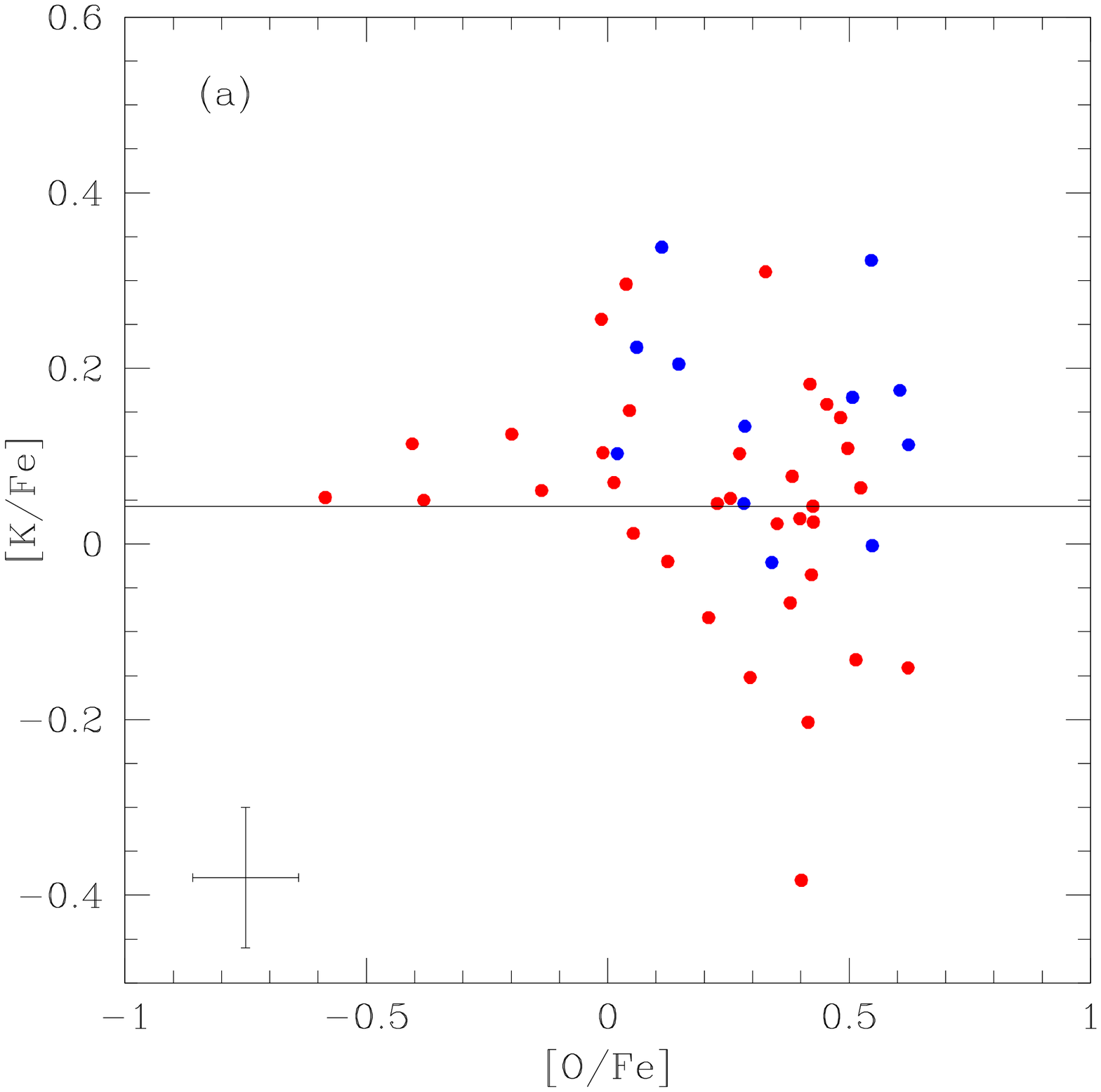}\includegraphics[scale=0.30]{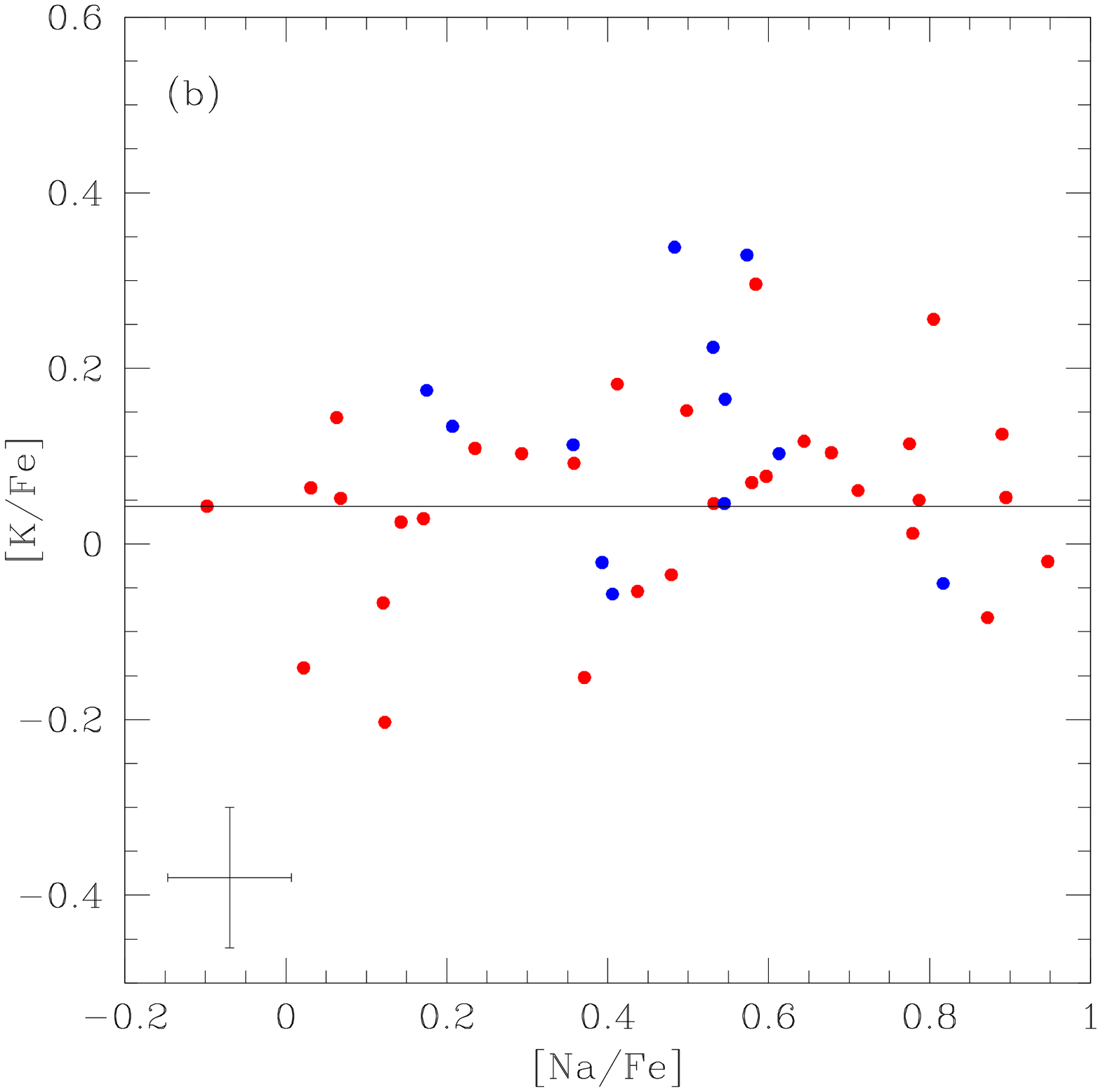}\includegraphics[scale=0.30]{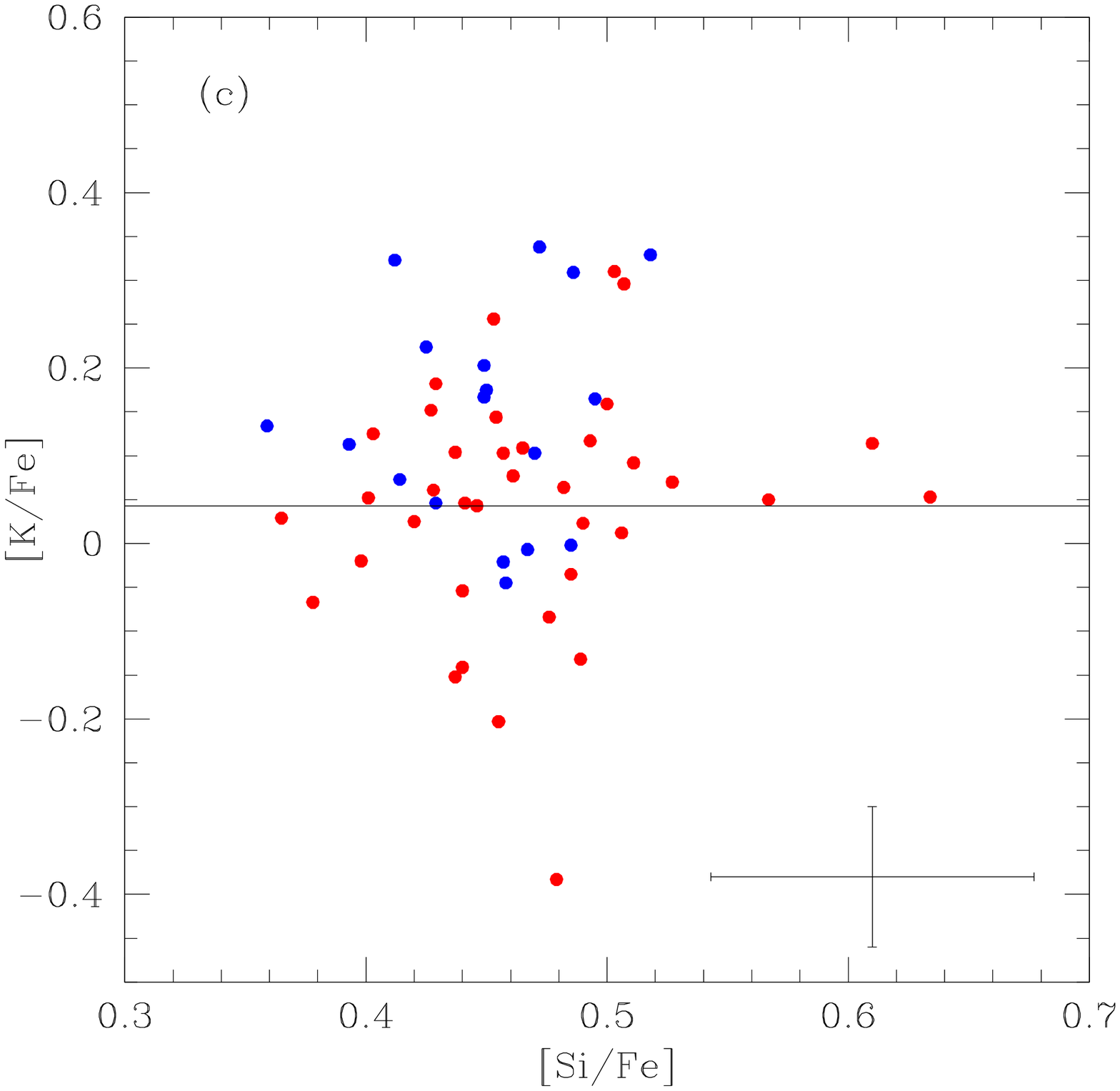}
\caption{Comparison of K abundances to the abundance ratios [O/Fe] (left panel),
[Na/Fe] (middle panel), and [Si/Fe] (right panel). Symbols and error bars are
the same  as in Fig.~\ref{f:fig3}.}
\label{f:fig5}
\end{figure*}

\begin{figure*}
\centering
\includegraphics[scale=0.40]{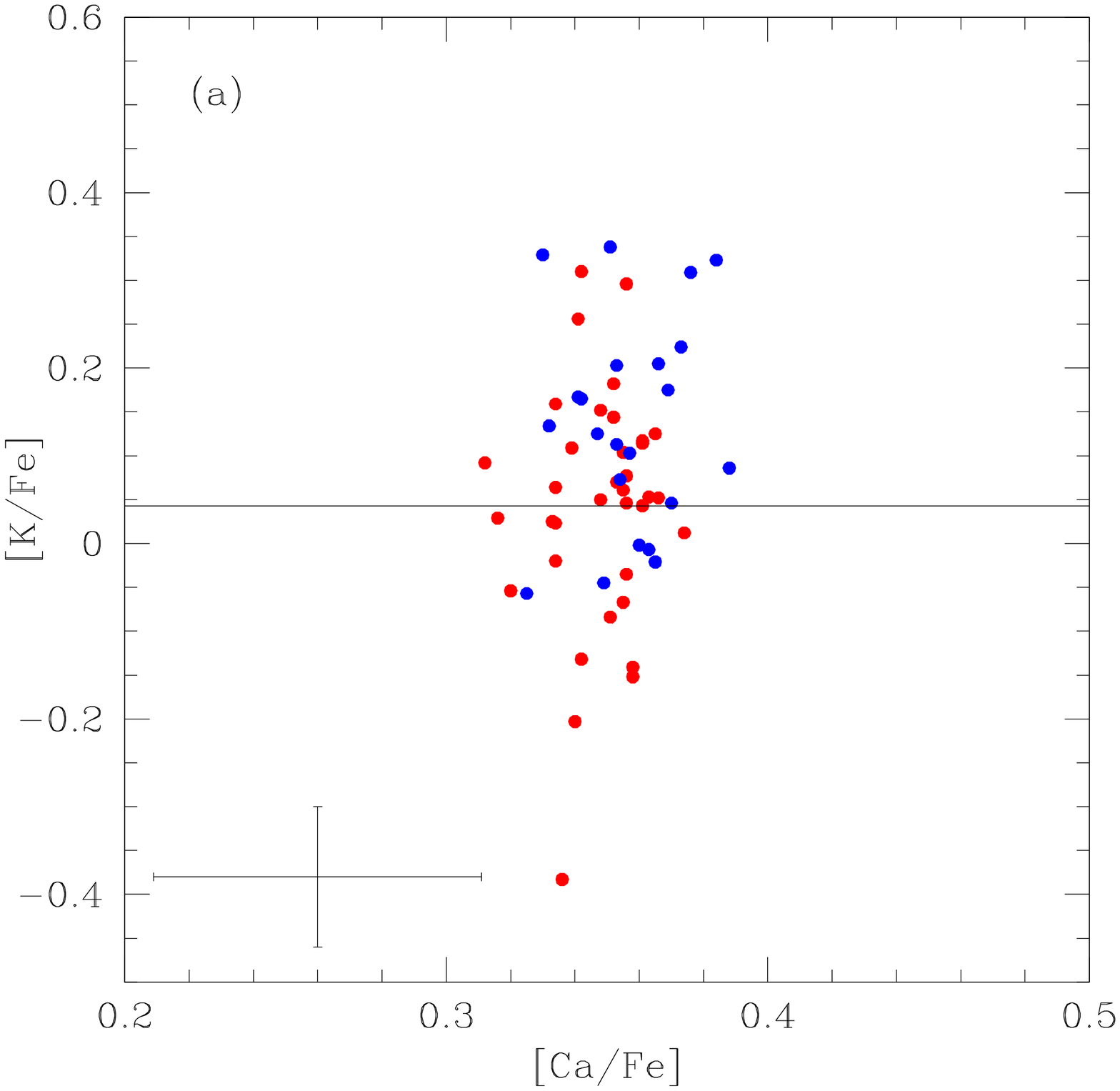}\includegraphics[scale=0.40]{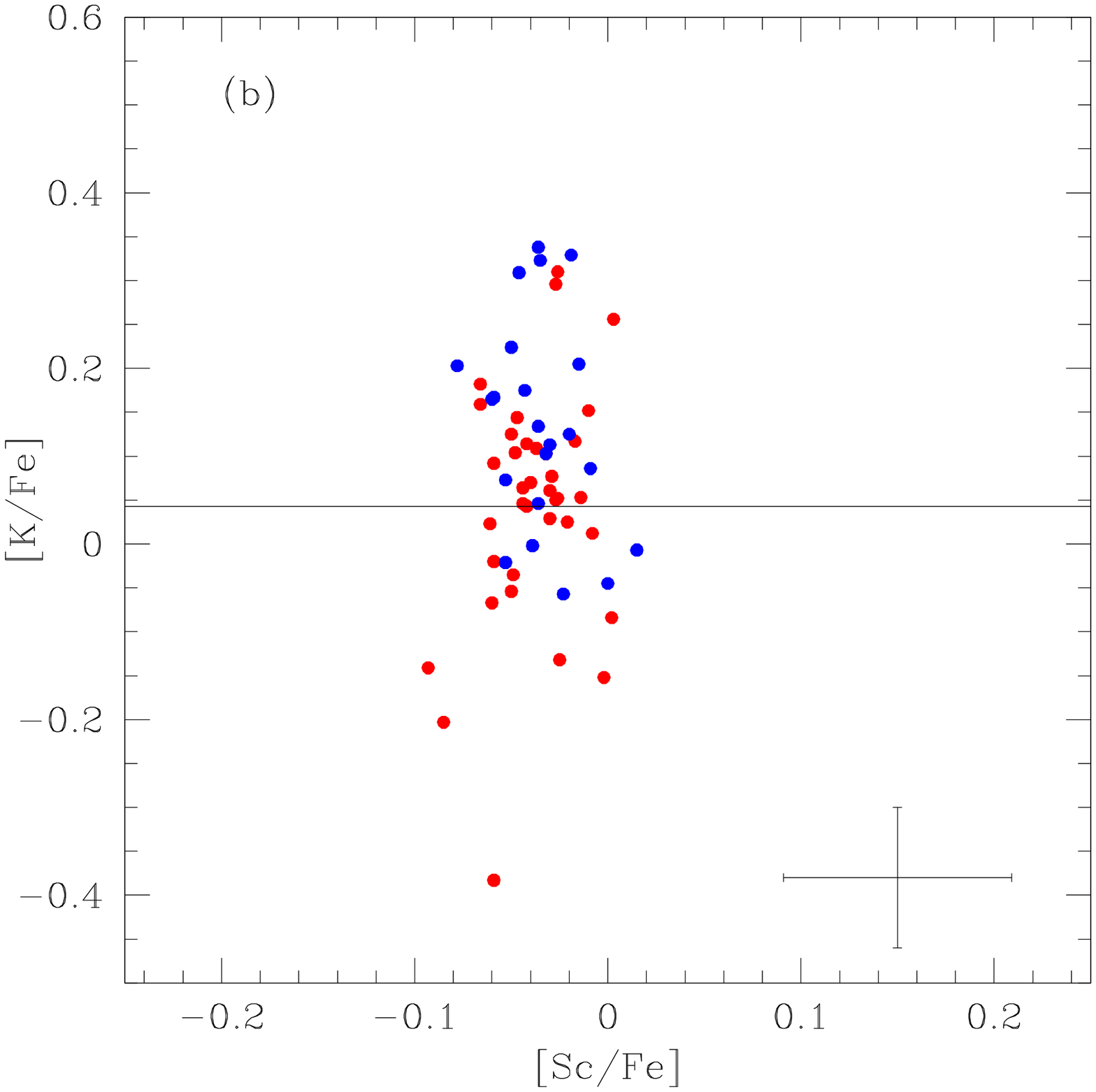}
\caption{Comparison of K abundances to the abundance ratios [Ca/Fe] (left 
panel) and [Sc/Fe] (right panel). Symbols and error bars are  the same as in
Fig.~\ref{f:fig3}.}
\label{f:fig6}
\end{figure*}

The same conclusion is reached from Fig.~\ref{f:fig4}, where we plotted the
cumulative distributions of the [K/Fe] ratios for the higher and lower Mg samples.
The null hypothesis (the two distributions are extracted from the same parent
population) can be safely rejected by a Kolmogorov-Smirnov test ($p=0.021$). 
If only stars with high S/N ($>100$) spectra are considered, this
probability decreases to $8.8\times 10^{-3}$, and the difference of the mean
values for [K/Fe] in the two Mg groups is even more significant 
($p=2.7\times 10^{-3}$).

Only 38 stars out of the 59 giants with derived K abundances also have [Mg/Fe] 
obtained in Carretta et al. (2014). This number increases to 46, 44, 55, 59, and
59 if we consider O, Na, Si, Ca, and Sc, providing a good opportunity to perform
a consistency test on the astrophysical nature of abundance variations of K for
NGC~4833.
If the spread observed for K is due to the same network of
proton-capture reactions responsible for shaping the chemical signature of
multiple populations, and since the potassium abundances are anti-correlated to
those of Mg, we should expect the [K/Fe] ratio to be correlated to light
elements produced in these reactions and anti-correlated to those depleted in
this H-burning.

This occurrence is examined in Fig.~\ref{f:fig5}, where the abundances of K 
are compared to those of O, Na, and Si from Carretta et al. (2014). The [K/Fe]
ratio decreases as O abundances increase, whereas it is roughly correlated to Na
and Si abundances.
 
Finally, the run of [K/Fe] values as a function of abundances of Ca and Sc is
shown in the two panels of Fig.~\ref{f:fig6}. In these panels, the x-axis scale
has been enlarged to better show the small variations seen in Ca and Sc
abundances. We remind readers that the r.m.s. scatter found for K is more than six times
the r.m.s. for the other two elements (about 0.02 dex from GIRAFFE spectra).
Nevertheless, K abundances seem to be correlated to small star-to-star
variations in both Ca and Sc, whose excesses with respect to the level of
unpolluted field stars were clearly highlighted in Carretta and Bragaglia
(2021) for NGC~4833.

\subsection{Comparison with Roederer and Thompson (2015)}

\begin{figure}
\centering
\includegraphics[scale=0.50]{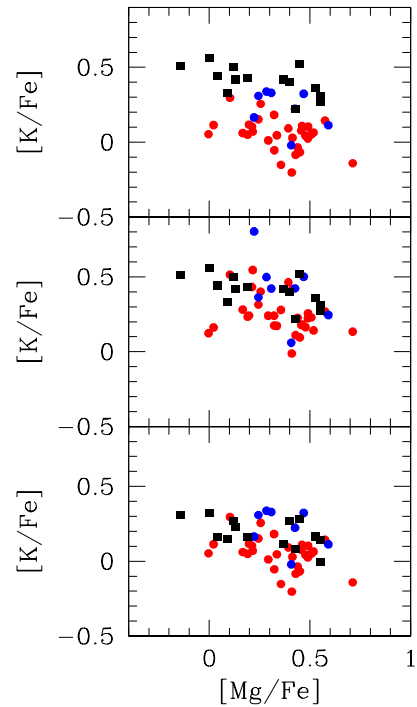}
\caption{[K/Fe] ratios as a function of [Mg/Fe] ratios in NGC~4833. Upper panel:
Our data are compared to the original data from Roederer and Thompson (2015:
black filled squares). Middle panel: All K abundances were derived using the
microturbulent velocity $v_t$ derived from Fe~{\sc i} lines. Lower panel: All K
abundances were obtained using $v_t$ values from the relation as a function of
surface gravity.}
\label{f:figroe}
\end{figure}

Roederer and Thompson (2015) analysed several elements in 15 giants of NGC~4833.
As they discussed, differences in the derived abundances with respect to 
Carretta et al. (2014) are explained by the different scale of atmospheric
parameters and solar reference abundance. However, when comparing abundances of
potassium, we found a clear offset between the present analysis and their work
(Fig.~\ref{f:figroe}, upper panel, black squares).

We examined the possible causes for this discrepancy. The offset cannot be due
to the adopted solar abundances for K and Fe, which imply a shift by 0.05 dex
only. Corrections for non-LTE effects were derived from Takeda et al. (2002) in
both studies. Roederer and Thompson (2015) used the K~{\sc i} 7664.87~\AA\ line,
which is not contaminated in their spectra, but after correcting for non-LTE they
found that the two lines give the same abundance within 0.02 dex.

The reason is very likely due to the adopted values of the microturbulent
velocity. Roederer and Thompson (2015) adopted the values of $v_t$ derived from
Fe~{\sc i} lines, as is done in all the standard abundance analyses. As a
consequence, we found that their K abundances show a trend as a function of
$v_t$. In the middle panel of Fig.~\ref{f:figroe}, we use for our 59 stars our
first analysis, so that all K abundances shown in the panel were derived using
$v_t$ values from Fe~{\sc i} lines. There is no offset anymore. As a further
confirmation, in the lower panel of Fig.~\ref{f:figroe} all abundances of K were
obtained using the relation from Worley et al. (2013) for $v_t$ as a function of
the surface gravity. The [K/Fe] ratios from the present work are in excellent
agreement with those by Roederer and Thompson (2015), which are consistently
anti-correlated to the [Mg/Fe] ratios.

\section{NGC~4833 in context}

The results of the present analysis both extend and confirm previous findings of
a Mg-K anti-correlation in NGC~4833. These star-to-star abundance variations are
not as extreme as those found in NGC~2419 (Cohen and Kirby 2012, 
Mucciarelli et al. 2012): NGC~4833 is more similar to NGC~2808, despite the
large difference in metallicity.

\begin{figure*}
\centering
\includegraphics[scale=0.30]{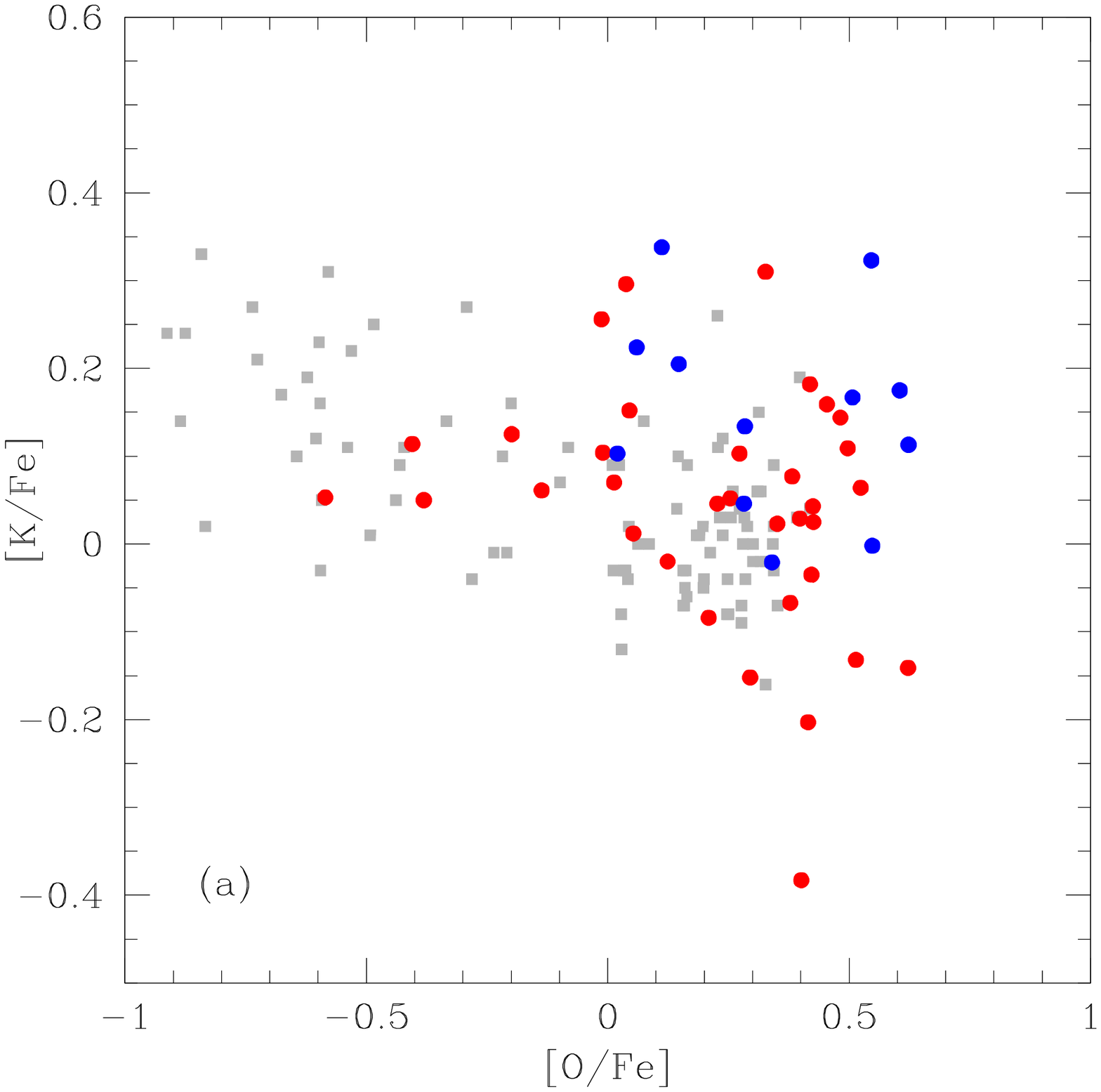}\includegraphics[scale=0.30]{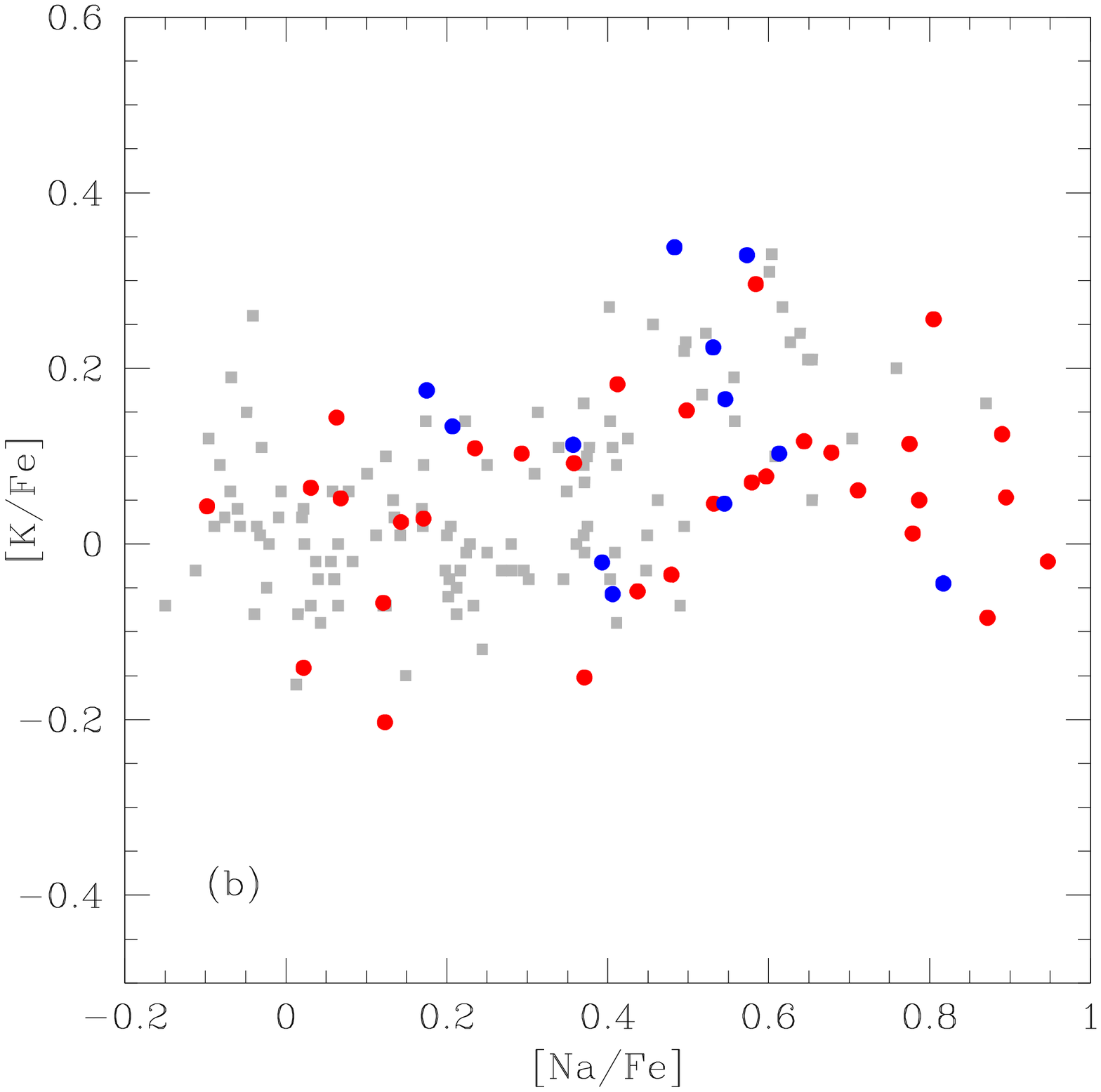}
\includegraphics[scale=0.30]{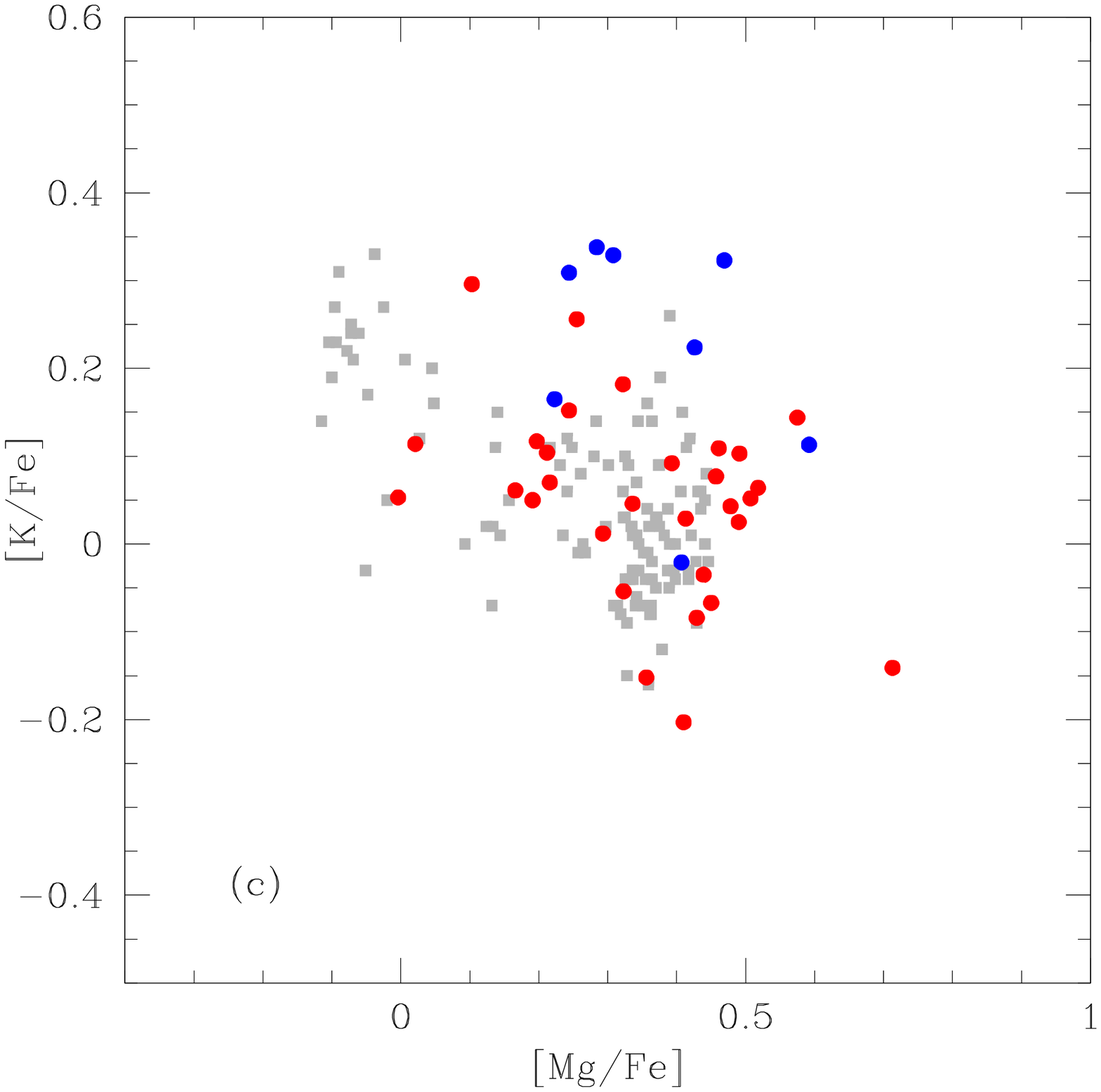}\includegraphics[scale=0.30]{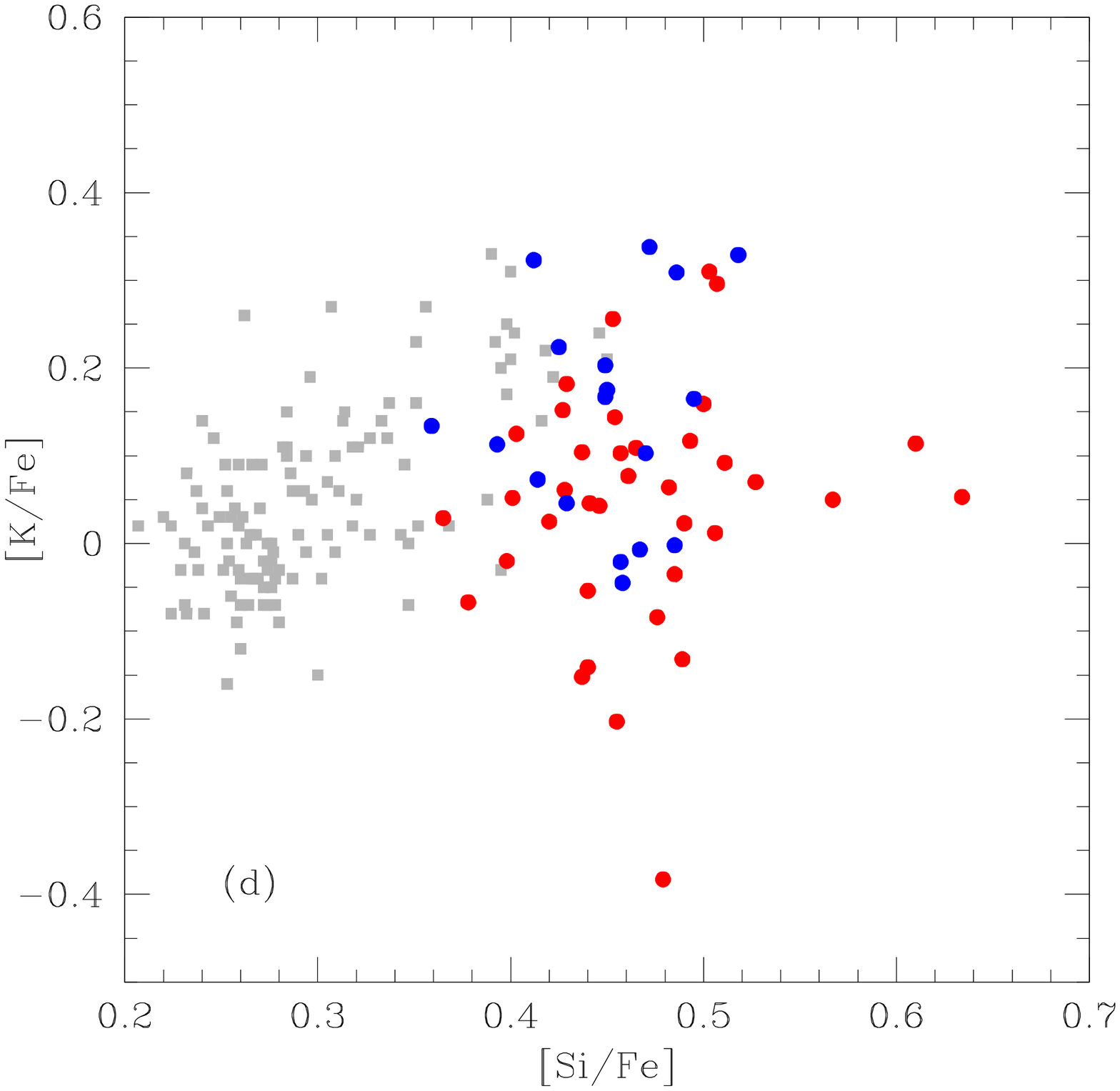}
\includegraphics[scale=0.30]{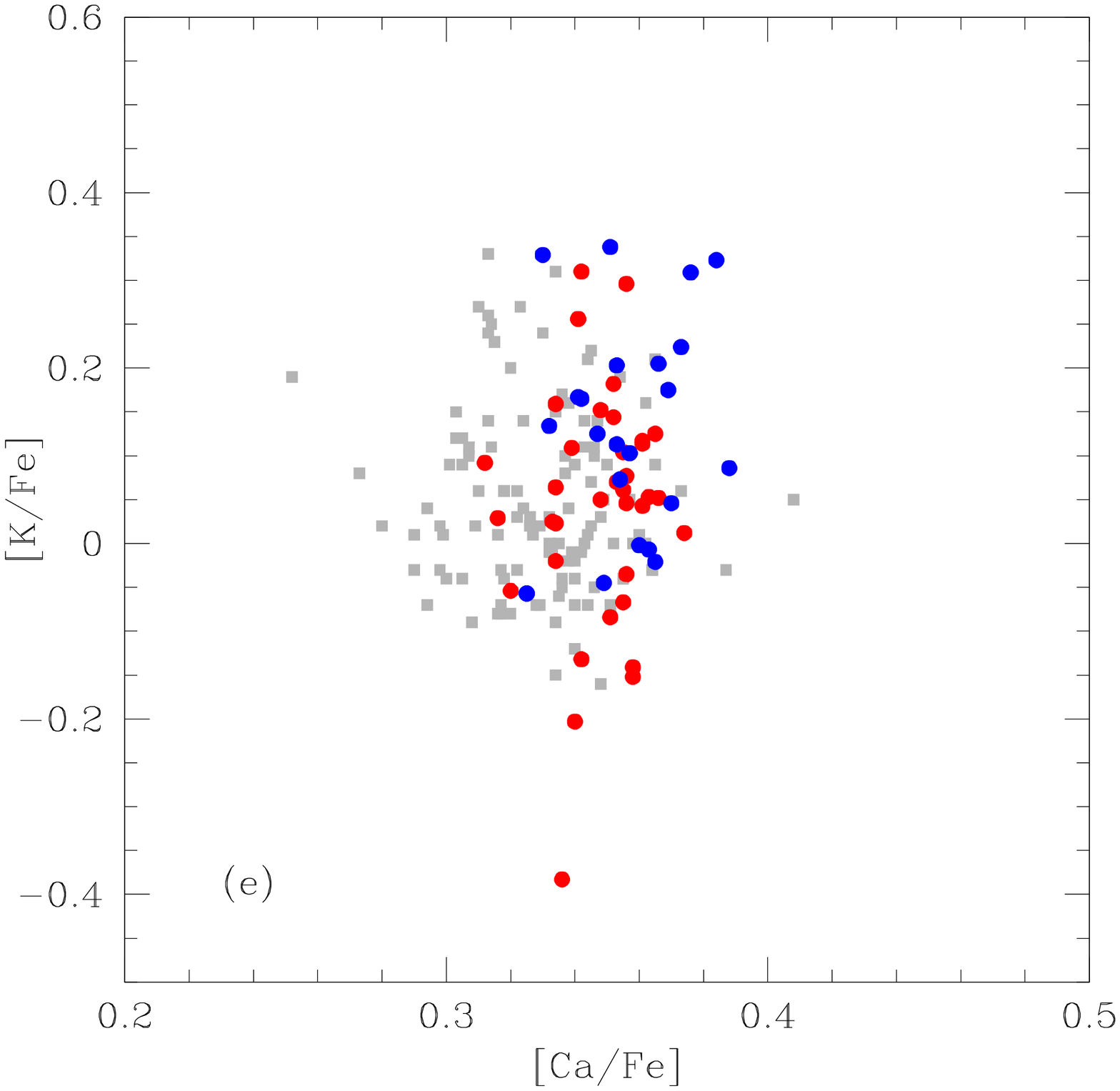}\includegraphics[scale=0.30]{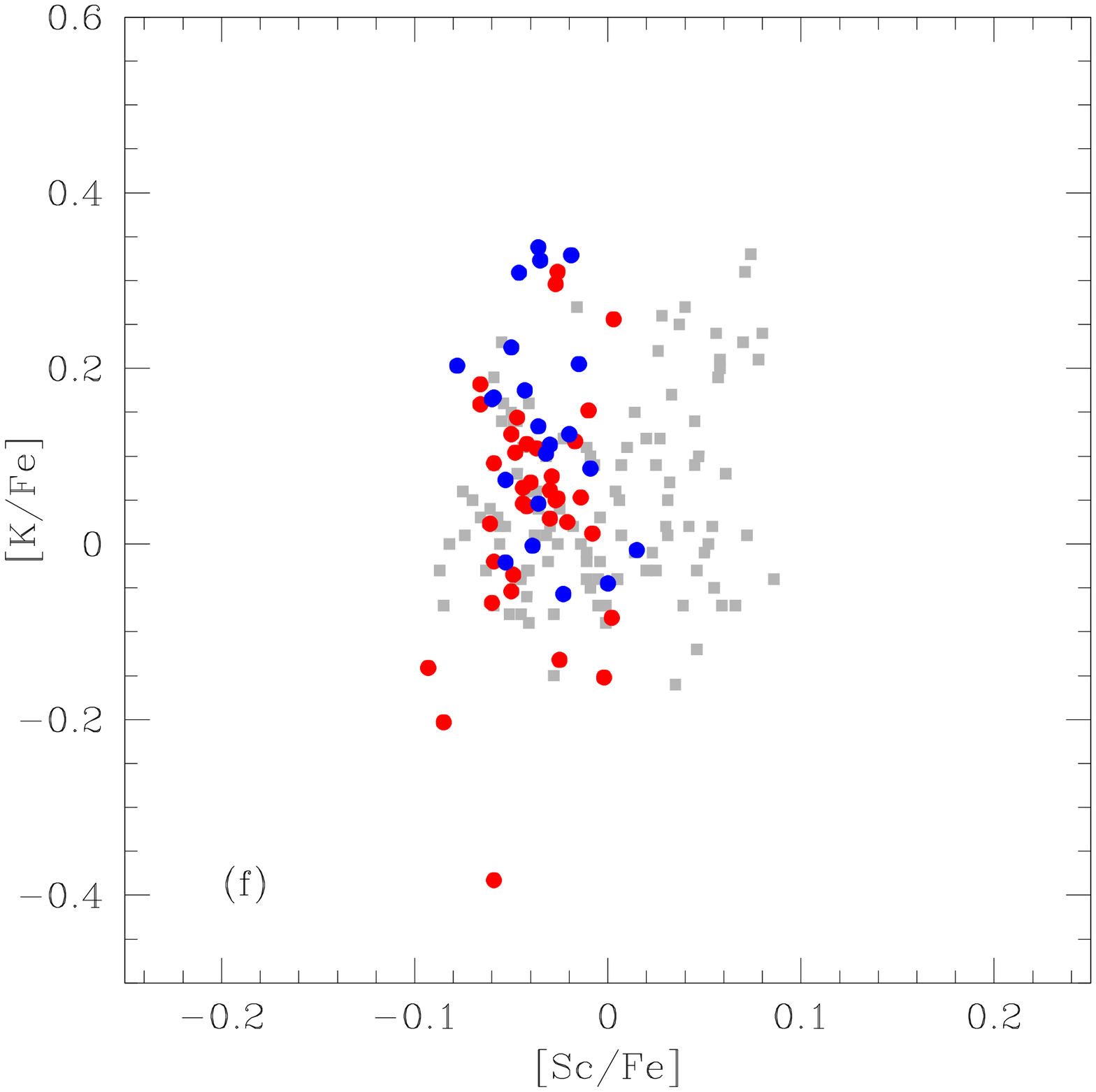}
\caption{Abundance ratios [K/Fe] in NGC~4833 and NGC~2808 as a function of
abundances of O, Na, Mg, Si, Ca, and Sc. For NGC~4833, the symbols are the same as in
Fig.~\ref{f:fig3}. Stars in NGC~2808 are indicated by grey squares.}
\label{f:figtot}
\end{figure*}

In Fig.~\ref{f:figtot} we compare the pattern of light elements in NGC~4833 
(present analysis and Carretta et al. 2014) and in NGC~2808 (Carretta
2015 and Mucciarelli et al. 2015). Small residual offsets may exist concerning
K abundances in NGC~2808 because Mucciarelli et al. (2015) adopted temperatures
from Carretta et al. (2006), whereas in Carretta (2015) stars in NGC~2808 were 
analysed with a slightly different scale, but in general the abundance scale is
as homogeneous as possible in the two GCs.

Apart from the clear offset in the average level of silicon, abundances in
NGC~4833 trace those in NGC~2808 rather well, although on a smaller scale. The
abundance spreads in O, Mg, Ca, and Sc observed in NGC~2808 are larger than the
equivalent in NGC~4833, but the overall pattern is very similar.
In both GCs, stars with low Mg abundances (reaching down to sub-solar
values in a good fraction of stars in NGC~2808) tend to show high K abundances.

\begin{figure}
\centering
\includegraphics[scale=0.40]{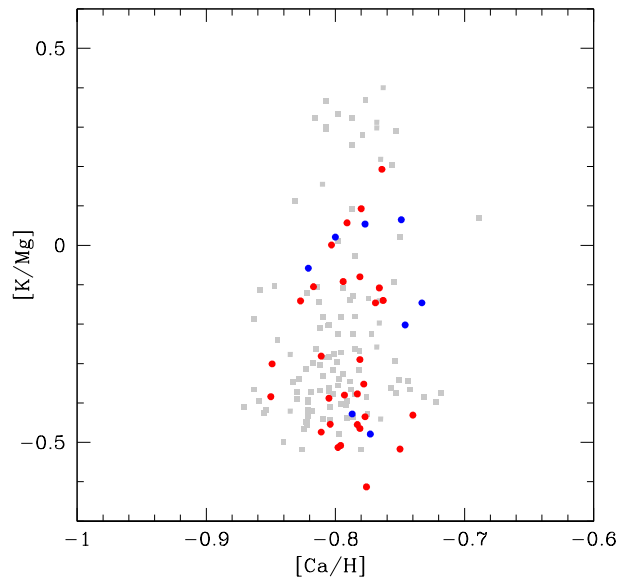}
\caption{[K/Mg] ratios as a function of the [Ca/H] ratios in NGC~4833 (same symbols
as in Fig.~\ref{f:fig3}) and in NGC~2808 (grey squares). An offset of 0.9 dex
was added to the [Ca/H] ratios in NGC~4833 to account for the difference in
metallicity with respect to NGC~2808.}
\label{f:summary}
\end{figure}

A compact summary of the relation among K, Mg, and Ca is given in
Fig.~\ref{f:summary}, where the ratio [K/Mg] is plotted as a function of [Ca/H]
in these two GCs. The values of [Ca/H] in NGC~4833 were shifted by applying an
offset of 0.9 dex to account for the difference in metallicity between the two
clusters. Again, apart for a larger spread observed in NGC~2808, there is a good
match for the abundances of these species.  The giants with higher abundances of
K also tend to be slighly more rich in Ca (and Sc) in both NGC~2808 and
NGC~4833. Both clusters are among the eight GCs where Carretta and Bragaglia (2021)
recently found significant excesses of both species with respect to the level of
field stars with a similar metallicity.

The present work seems to confirm that whenever significant star-to-star
variations in Ca and Sc are detected, they are found to be accompanied by even
larger variations in the K content. Evidence for this was provided
by Cohen and Kirby (2012) and Carretta et al. (2013) in NGC~2419, by Carretta
(2015) and Mucciarelli et al. (2015) in NGC~2808, and finally in NGC~4833 here.
Also spectroscopic infrared data from APOGEE  agree with this scenario nicely
since the anti-correlation between K and Mg abundances in $\omega$ Cen
(M\'esz\'aros et al. 2020) is paralleled to statistically significant variations
in Ca and Sc (Carretta and Bragaglia 2021).
Interestingly enough, Carretta and Bragaglia (2021) did not instead find robust
evidence of Ca or Sc excesses in NGC~1904, a cluster whose K-Mg anti-correlation
was argued by M\'esz\'aros et al. (2020) to be probably only due to the large
uncertainties associated with the derived abundances.

\begin{figure*}
\centering
\includegraphics[scale=0.30]{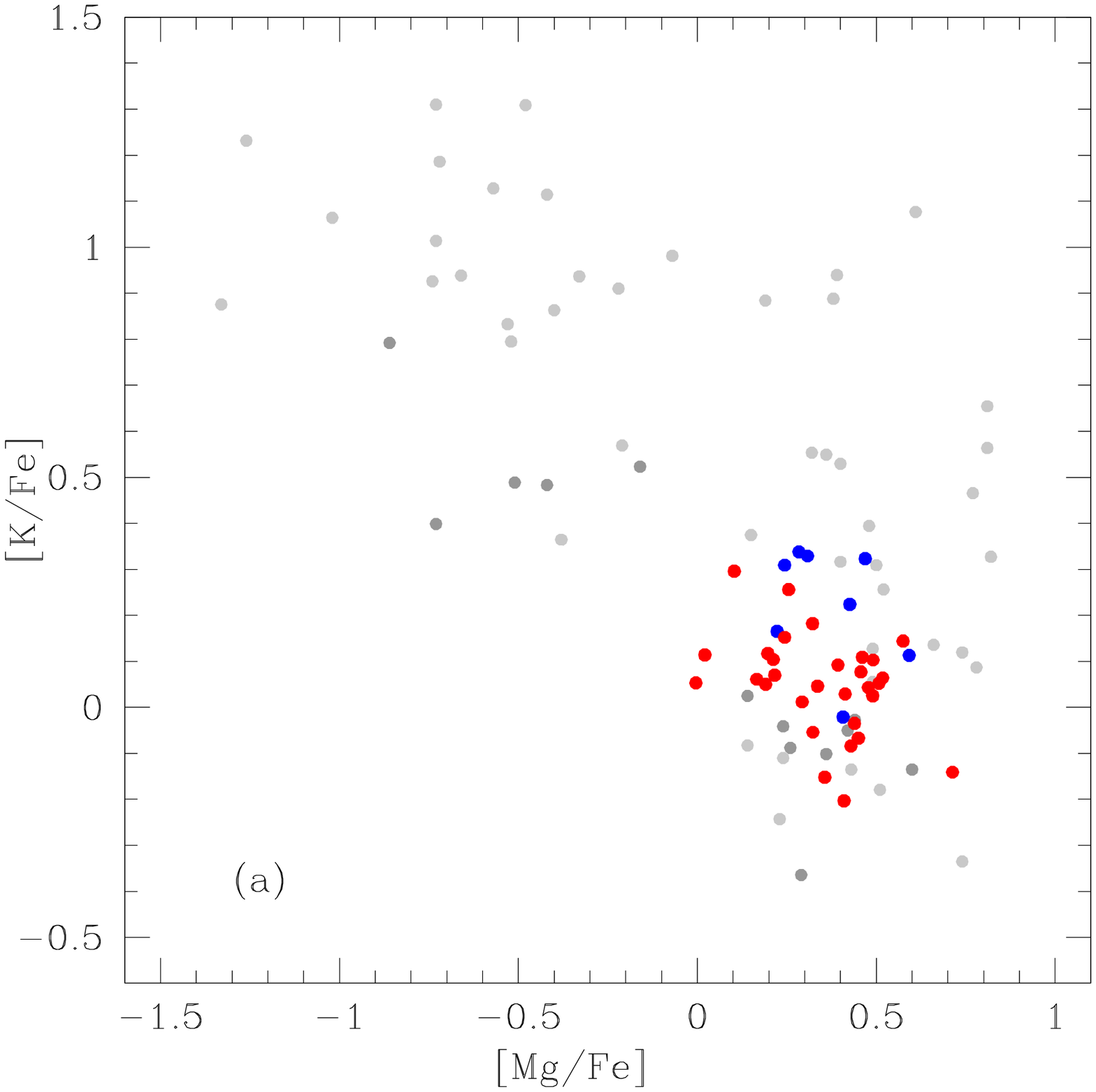}\includegraphics[scale=0.30]{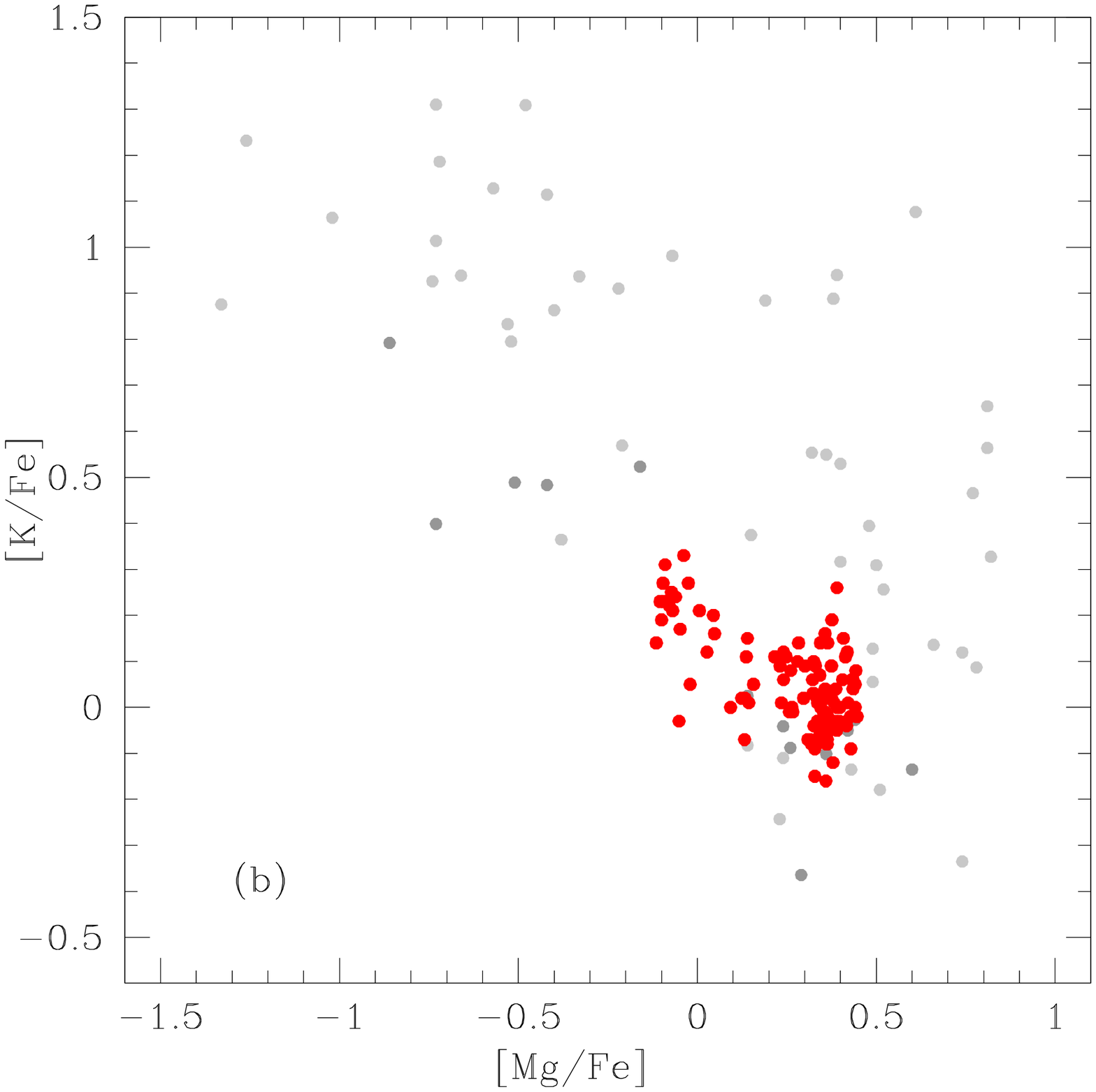}\includegraphics[scale=0.30]{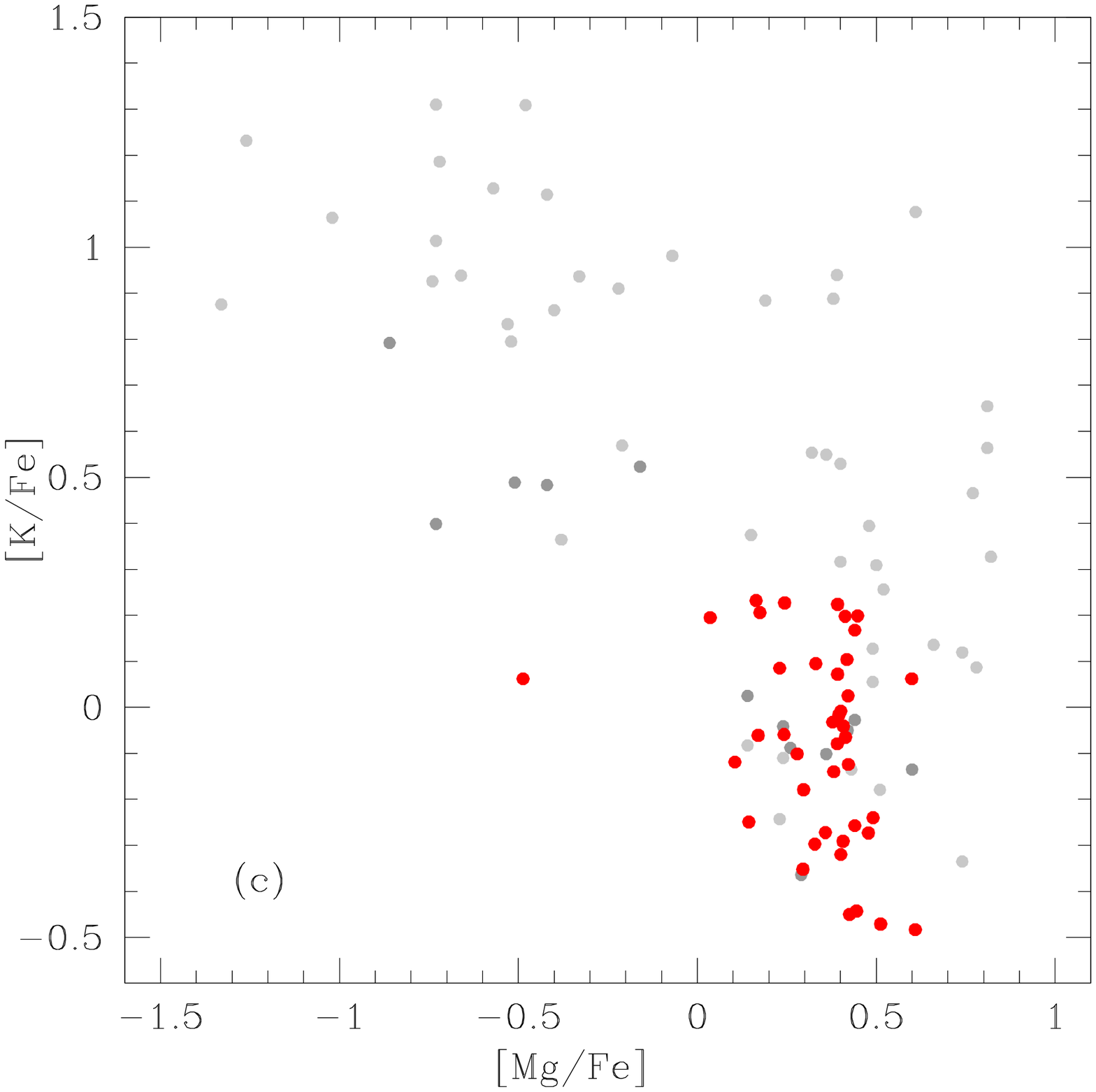}
\caption{Comparison of the K-Mg anti-correlation in different GCs. Left panel:
Red and blue circles are the stars of the present work (same symbols as in
Fig.~\ref{f:fig3}) for NGC~4833 superimposed to stars in NGC~2419 from Cohen and
Kirby (2012: dark grey circles) and Mucciarelli et al. (2012: light grey
symbols). Middle panel: Red circles are stars in NGC~2808 from Carretta (2015)
and Mucciarelli et al. (2015). Right panel: Red circles indicate stars in
NGC~6715 (M~54) from the preliminary analysis by Carretta (in prep.).}
\label{f:cfr2419}
\end{figure*}

To put data for NGC~4833 in context, in Fig.~\ref{f:cfr2419} we compare the
results from the present work as well as for NGC~2808 to the K-Mg
anti-correlation observed in NGC~2419 (left and middle panels, respectively).
From these plots, it is clear again that depletions in Mg are more extreme in
NGC~2808 than in NGC~4833, whereas the range spanned by K is more or less the
same. Moreover, the large difference with respect to
the case of NGC~2419 is immediately evident. All the variations in K content for the sub-populations of
stars with primordial, intermediate, and extreme composition (see Carretta et
al. 2009a for a definition of these different stellar populations in
GCs) are confined in the same region of the K-Mg plane in normal GCs. However,
even the stars with the most enhanced K abundances in these GCs are restricted in 
the group of stars with solar [K/Fe] ratios, on average, in NGC~2419.
This peculiar GC also shows a noticeable fraction of giants with much more
extreme values of K and Mg that have no correspondence in normal GCs: The
chemical inventory of NGC~2419 appears to be really unique among GCs.

A possibility to explain this uniqueness was discussed in Carretta et al.
(2013), where we argued that the extreme population could be due to the
polluting contribution of a pair-instability supernova (PISNe). The extreme
rarity of such a peculiar SN (Ren et al. 2012) would account for the fact that
the extreme population is only found (at least so far) in one cluster out of the
about 150 of the whole GC population in the Milky Way. 
The strong overabundance of Ca going with extreme Mg depletions, predicted for
polluting matter provided by PISN, was illustrated in Carretta et al. (2013) by
using the [K/Mg] ratio as a function of [Ca/H]. For comparison, in NGC~2419 the
[K/Mg] values of the extreme population would be off scale in
Fig.~\ref{f:summary}, reaching the 2 dex level in the data by Mucciarelli et al.
(2012).

Other predictions for PISNe (such as a strong odd-even effect in the chemical
inventory) are not satisfied, however (see the detailed discussion in Carretta
et al. 2013). At present, we can only reiterate that neither NGC~4833 nor
NGC~2808 belong to the same class of NGC~2419.

\begin{figure}
\centering
\includegraphics[scale=0.40]{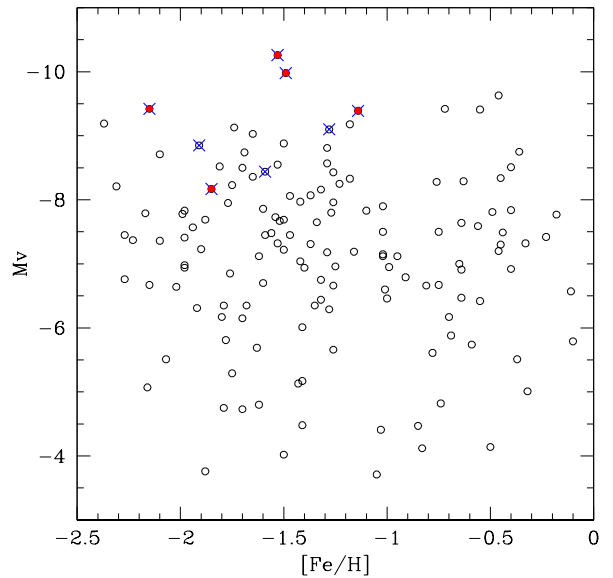}
\caption{Total absolute magnitude of GCs in the Milky Way (from Harris 1996,
2010 online edition) as a function of metallicity [Fe/H]. Blue crosses indicate
GCs with significant excesses of Ca and Sc (Carretta and Bragaglia 2021). The
filled red circles are superimposed to GCs, also showing significant variations
in K abundances (see text).}
\label{f:mvfe}
\end{figure}

In Fig.~\ref{f:mvfe}, we plotted the total absolute magnitude $M_V$ (a proxy for the
observed, present-day mass of clusters) for the GCs in the Milky Way as a
function of the metal abundance [Fe/H], from the Harris (1996, 2010 online
edition) catalogue. Blue crosses indicate the eight GCs where Carretta and Bragaglia
(2021) found significant excesses of Ca and Sc with respect to field stars.
In order of increasing metallicity the GCs are NGC~2419, NGC~5824, NGC~4833,
NGC~5139 ($\omega$ Cen), NGC~5986, NGC~6715 (M~54), NGC~6402, and NGC~2808.
Red filled circles are superimposed to GCs where significant variations in K
were detected: NGC~2419, NGC~2808, NGC~5139, and NGC~4833.

To these GCs, we also tentatively added NGC~6715 because preliminary results
(Carretta, in preparation) of an
analysis analogue to the present study seem to reveal
a large scatter in K abundances, which are possibly anti-correlated to the Mg
abundances from Carretta et al. (2010a), as shown in the right panel in
Fig.~\ref{f:cfr2419}. We caution, however, that this is still a work in
progress and should be taken cum grano salis because the quality of
spectra is lower than those used for NGC~4833 and even using the relation by
Worley et al. (2013) the trend of K abundances as a function of $v_t$ is not
completely eliminated, probably owing to the relatively large intrinsic
dispersion in iron abundances in this cluster ($\sim 0.19$ dex, Carretta et al.
2010a).

The GCs where the effect of alterations of the abundances of K, Ca, and Sc is
more evident appear to be confined to a region identified by high mass and low
metallicity (Fig.~\ref{f:mvfe}). In the observational landascape associated with
chemical signatures of multiple stellar populations in GCs, this is not
unexpected. 
We have known for a long time that the minimum observed ratio [O/Fe]$_{\rm min}$ in GCs
is well reproduced by a linear bivariate relation as a function of cluster
luminosity and metallicity (Carretta et al. 2009a). Analogously, the amount of
the [Al/Fe]$_{\rm prod}$ produced is found to be a function of a linear
combination of the same parameters (Carretta et al. 2009b). Finally, the
fraction of outliers with large excesses of Ca and Sc is again found to be well
reproduced by a combination of $M_V$ and [Fe/H].

These findings, together with the sort of segregation observed in
Fig.~\ref{f:mvfe}, prompted us to verify if the variations in K may also depend
on these global cluster parameters.
To better quantify the observed spread in K, we used the interquartile range of
the [K/Fe] ratio, IQR[K/Fe]. By definition, the IQR is less sensitive to the
outliers and, as shown for the [O/Na] ratio by Carretta (2006) and Carretta et
al. (2010b), is a robust quantitative measure for the extent of the nuclear
processing affecting polluted populations in GCs.

\setcounter{table}{2}
\begin{table}
\centering
\caption{Interquartile range IQR[K/Fe], luminosity, and metallicity for
selected GCs.}
\begin{tabular}{lccrlc}
\hline
GC  & IQR     & ref.   & $M_V$ & [Fe/H] & ref.   \\
    & [K/Fe]  & [K/Fe] &       &        & [Fe/H] \\
\hline
NGC~0104 & 0.110 & a & $-$9.42 & $-$0.768 & e \\
NGC~2419 & 0.634 & b & $-$9.42 & $-$2.15  & f \\
NGC~2808 & 0.140 & c & $-$9.39 & $-$1.151 & g \\
NGC~4833 & 0.151 & d & $-$8.17 & $-$2.015 & h \\
NGC~6752 & 0.128 & a & $-$7.73 & $-$1.555 & i \\
NGC~6809 & 0.130 & a & $-$7.57 & $-$1.934 & e \\
\hline
\end{tabular}
\begin{list}{}{}
\item[-] References for observed K abundances: a=Mucciarelli et al. (2017);
b=average from Cohen and Kirby (2012) and Mucciarelli et al. (2012);
c=Mucciarelli et al. (2015); and d=this work.
\item[-] References for metallicity: e=Carretta et al. (2009c); f=Harris (1996, 2010
online edition); g=Carretta (2015); h=Carretta et al. (2014); and i=Carretta et al.
(2007b).
\item[-] Total absolute magnitudes are from Harris (1996, 2010 online edition).
\end{list}
\label{t:tab3}
\end{table}

The required parameters are listed in Table~\ref{t:tab3} for six selected GCs.
From the sample, we had to exclude the two most massive GCs in the Galaxy, namely
$\omega$ Cen and M~54, for several reasons. Unlike the others, they are not 
mono-metallic GCs, showing a large spread in iron, and they both are presumably
the nuclear remnants of ancient dwarf galaxies that merged with the Milky Way (e.g.
Bekki and Freeman 2003, Bellazzini et al. 2008). The abundances of K derived for
M~54 are still very preliminary, and those for $\omega$ Cen are obtained from IR
spectra in APOGEE, so that they are not homogeneous with the set of
abundances for the other GCs. Finally, the IQR[K/Fe] for $\omega$ Cen
significantly depends on the selection criteria adopted by M\'esz\'aros et al.
(2020) to extract the best sample. For the K-Mg anti-correlation, they used only
stars with a metallicity of [Fe/H]$<-1.5$ dex (their Figure 10), whereas it is well
assessed that in this GC the most extreme processing due to the multiple
population phenomenon is observed at intermediate metallicity, higher than -1.5
dex (e.g. Johnson and Pilachowski 2010, Marino et al. 2011). We then decided to
conservatively exclude these two GCs, although we point out that the results of
this check would not change with their inclusion in the sample.

\begin{figure*}
\centering
\includegraphics[scale=0.30]{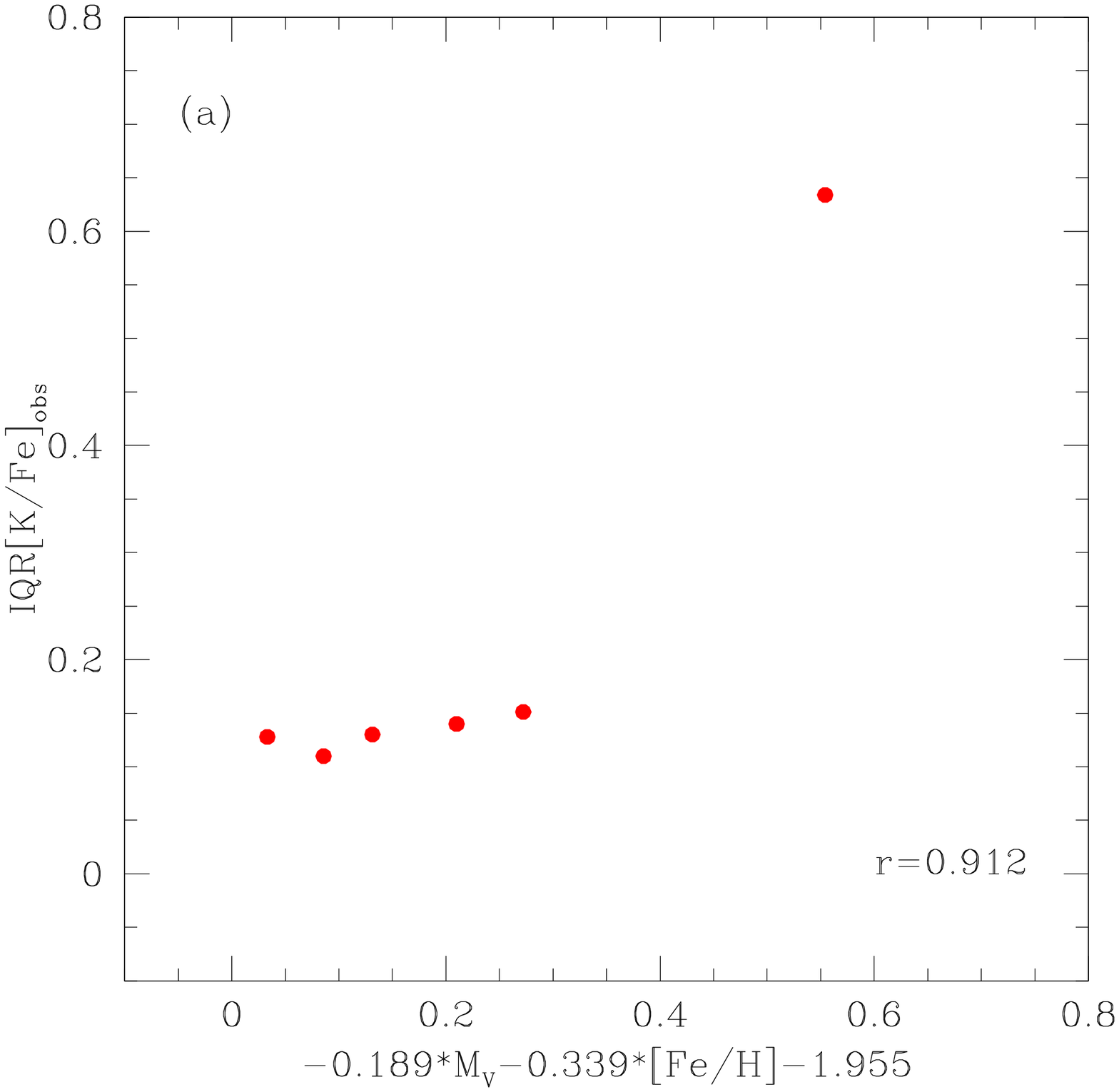}\includegraphics[scale=0.30]{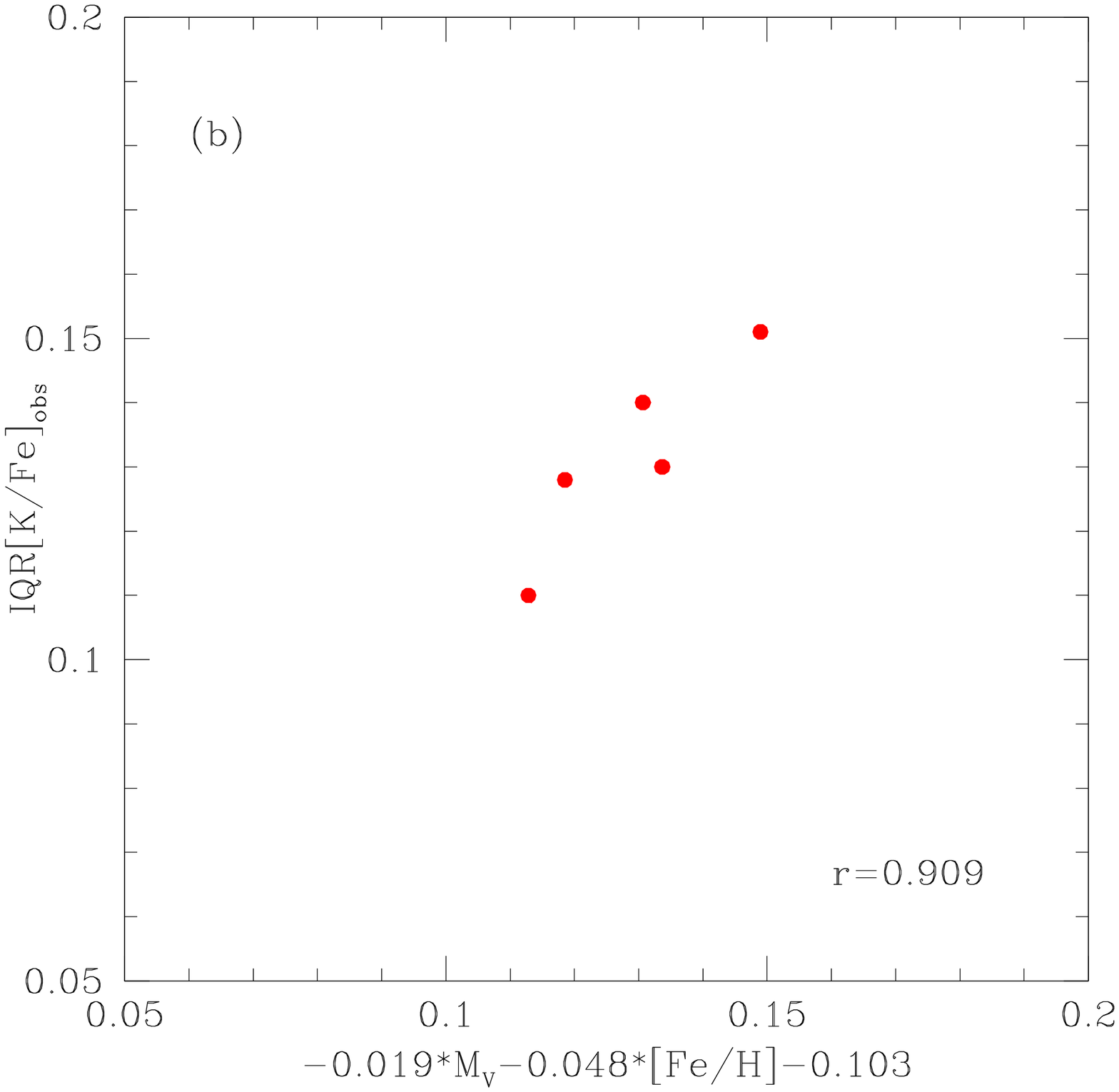}
\caption{Observed interquartile range of the [K/Fe] ratio (IQR[K/Fe]) as a
function of a bivariate linear combination of $M_V$ and [Fe/H] for the GCs in
Table~\ref{t:tab3}. In the left and right panel, NGC~2419 is included and
excluded from the fit, respectively.}
\label{f:bivar}
\end{figure*}

In the left panel of Fig.~\ref{f:bivar}, the observed values of IQR[K/Fe] are
plotted as a function of the bivariate combination of $M_V$ and [Fe/H]. The
Pearson correlation coefficient is also listed and obviously indicates that the
relation is highly significant. It is also obvious that NGC~2419 is
driving this correlation due to its extreme value. However, even when this
peculiar GC is excluded, the combined dependence on the cluster luminosity and
metallicity is very significant (probability $p=3.2\times 10^{-2}$) for the
other five normal GCs.

Admittedly, the sample is small because the set of GCs with large enough
numbers of
stars with derived K abundances is still very limited. However, the low probability
of obtaining such a statistically tight relation by mere chance seems to suggest
that the extent of star-to-star variations in the K content is also related to
the global cluster parameters as are other quantities (minimum oxygen reached,
aluminium production, excesses of Ca and Sc). The conclusion is once again that
in very massive and (or) metal-poor clusters, the K abundance variations are a
good indicator of the H-burning proton capture reactions at very high
temperatures.

\section{Summary}

We studied the distribution of potassium abundances in the metal-poor globular
cluster NGC~4833, whose multiple stellar populations show a rather extreme
chemical composition. 
We measured [K/Fe] ratios using the high resolution GIRAFFE setup HR18, whose
spectral range includes the K~{\sc i} resonance doublet. The abundances of 
K were derived for 59 stars from the line at 7698.98~\AA\, after cleaning the
spectra for a small contamination from telluric lines. The abundances were
corrected for non-LTE effects.
Stellar parameters and abundances of proton-capture elements were already
homogeneously obtained for the sample of stars in Carretta et al. (2014).

Our data show an anti-correlation between abundances of Mg and K, similar to the
one observed in NGC~2808 by Mucciarelli et al. (2015), but much reduced with
respect to the huge variations detected in NGC~2419 (Cohen and Kirby 2012,
Mucciarelli et al. 2012). 
In NGC~4833 we found that K abundances are correlated to elements enhanced in
proton-capture reactions (Na, Si) and anti-correlated to species depleted in the
same network of nuclear reactions (Mg, O). 

Evidence of slightly higher Ca and Sc abundances are observed among stars with
the highest K abundances. These findings suggest that in NGC~4833, the FG polluters
were acting to such a high temperature so as to also efficiently produce heavier
elements such as K and possibly Ca, confirming the recent observations by Carretta
and Bragaglia (2021) who detected statistically significant excesses of Ca and
Sc in the fraction of stars with low Mg abundances in this cluster.

In coupling the census of 77 GCs by Carretta and Bragaglia (2021) with the sample
of the few GCs where K abundances were extensively derived, we can appreciate how
significant star-to-star abundance variations in K are always accompanied by
smaller increases in Ca (and possibly Sc). The method ideated in Carretta and
Bragaglia (2021) then represents an advantageous asset, since it is  able to
pick up chemical patterns generated in processes of H-burning at a very high
temperature using abundances of more widely studied elements such as Ca and Sc, as
opposed to the paucity of data for K in GCs. 

We also found that the dispersion in K within an individual GC, measured by the
interquartile range of the [K/Fe] ratio, is well reproduced by a linear
combination of the cluster total absolute magnitude $M_V$ and metallicity
[Fe/H], as are other indicators of the most extreme processing in multiple
stellar populations. 
Although the sample of GCs used to derive this strong correlation is quite
limited, this discovery lends support to the scenario where the production of K
is due to proton-capture reactions occurring at very high temperatures ($>10^9$
K) in stars
where the efficiency of H-burning is enhanced in low-metallicity
environments. The S-AGB stars are then a viable candidate (Ventura et al. 2012).

Then, in this framework, the peculiar globular cluster NGC~2808 may represent 
the true anomaly, being the most metal-rich GC ([Fe/H]$=-1.14$ dex) where
significant K abundance variations have been detected so far.
Clearly, other surveys for K abundances in GCs would be highly welcome.

\begin{acknowledgements}
This paper is based on data obtained from the ESO Science Archive Facility under
request number 586933.
I wish to thank Angela Bragaglia for valuable help and useful discussions, and
all the people in the ESO Archive group, maintaining and supporting the ESO
archive, which undoubtly qualifies as one of the largest and most efficient
telescope in the world.
This research has made large use of the SIMBAD database (in
particular  Vizier), operated at CDS, Strasbourg, France, and of the NASA's
Astrophysical Data System.
\end{acknowledgements}

\end{document}